\documentstyle[10pt,epsf,epsfig,dp_delphititle]{dp_delphi}
\makeindex
\def\DpPaperGroup{EP}
\def\DpPaperRef{2001-029}
\def\DpDate{19 February 2001}
\def\DpAuthors{DELPHI Collaboration}
\def\DpSubmit{(Accepted by Physics Letters B)}
\def\DpTitle{{Measurement of the Mass and Width of the W Boson
in {\boldmath $ {e^+e^-} $} Collisions at { \boldmath $ \sqrt{s}  =   189~  \mathrm{GeV}$ }}}
\def\DpComment{ }
\def\DpEMail{ }

\begin{document}
\makeatletter
\newcount\@tempcntc
\def\@citex[#1]#2{\if@filesw\immediate\write\@auxout{\string\citation{#2}}\fi
  \@tempcnta\z@\@tempcntb\m@ne\def\@citea{}\@cite{\@for\@citeb:=#2\do
    {\@ifundefined
       {b@\@citeb}{\@citeo\@tempcntb\m@ne\@citea\def\@citea{,}{\bf ?}\@warning
       {Citation `\@citeb' on page \thepage \space undefined}}%
    {\setbox\z@\hbox{\global\@tempcntc0\csname b@\@citeb\endcsname\relax}%
     \ifnum\@tempcntc=\z@ \@citeo\@tempcntb\m@ne
       \@citea\def\@citea{,}\hbox{\csname b@\@citeb\endcsname}%
     \else
      \advance\@tempcntb\@ne
      \ifnum\@tempcntb=\@tempcntc
      \else\advance\@tempcntb\m@ne\@citeo
      \@tempcnta\@tempcntc\@tempcntb\@tempcntc\fi\fi}}\@citeo}{#1}}
\def\@citeo{\ifnum\@tempcnta>\@tempcntb\else\@citea\def\@citea{,}%
  \ifnum\@tempcnta=\@tempcntb\the\@tempcnta\else
   {\advance\@tempcnta\@ne\ifnum\@tempcnta=\@tempcntb \else \def\@citea{--}\fi
    \advance\@tempcnta\m@ne\the\@tempcnta\@citea\the\@tempcntb}\fi\fi}
 
\makeatother
\begin{titlepage}
\pagenumbering{roman}
\CERNpreprint{\DpPaperGroup}{\DpPaperRef} 
\date{{\small\DpDate}} 
\title{\DpTitle} 
\address{\DpAuthors} 
\begin{shortabs} 
\noindent
%
\noindent

\newcommand{\approxgt}{\raisebox{-.5ex}{\stackrel{>}{\sim}}}
\newcommand{\approxlt}{\raisebox{-.5ex}{\stackrel{<}{\sim}}}
\newcommand{\mean}[1]{{\left\langle #1 \right\rangle}}
\newcommand{\mod}[1]{{\left|#1\right|}}
\newcommand{\abs}[1]{{\left\|#1\right\|}}
\newcommand{\s}{{\sim}}
\newcommand{\ra}{{\rightarrow}}
\newcommand{\lra}{{\leftrightarrow}}
\newcommand{\longra}{{\longrightarrow}}
\newcommand{\mstrut}{\rule{0cm}{5ex}}
\newcommand{\ov}[1]{\overline{#1}}
\newcommand{\tc}[1]{{\mbox{\tiny #1}}}
\newcommand{\Real}{{\mathcal{R}e}}
\newcommand{\Imag}{{\mathcal{I}m}}


\newcommand{\bmW}{{\mathbf{W}}}
\newcommand{\bmF}{{\mathbf{F}}}
\newcommand{\bmX}{{\mathbf{X}}}
\newcommand{\bmI}{{\mathbf{I}}}
\newcommand{\bmalpha}{{\boldmath{\alpha}}}
\newcommand{\bmtheta}{{\mathbf{\theta}}}
\newcommand{\bmsigma}{{\mathbf{\sigma}}}
\newcommand{\bmlambda}{{\mathbf{\lambda}}}
\newcommand{\bmtau}{{\mathbf{\tau}}}
\newcommand{\bmdag}{{\mathbf{\dagger}}}
\newcommand{\Delx}{{\Delta x}}
\newcommand{\Dely}{{\Delta y}}
\newcommand{\rphi}{{r-\phi}}
\newcommand{\sz}{{s-z}}
\newcommand{\xy}{{x-y}}
\newcommand{\zo}{{z_0}}
\newcommand{\modzo}{{\left| z_0 \right|}}
\newcommand{\meanzo}{{\left\langle z_0 \right\rangle}}
\newcommand{\bphi}{{\bar\phi}}
\newcommand{\bthe}{{\bar\theta}} 

\newcommand{\chisq}{{{\chi}^{2}}}
\newcommand{\chindof}{{\chisq / \mathrm{dof}}}

\newcommand{\grpsuth}{{\mathrm{SU}(3)}}
\newcommand{\grpsutw}{{\mathrm{SU}(2)}}
\newcommand{\grpuone}{{\mathrm{U}(1)}}

\newcommand{\qq}{{q{\bar q}}}
\newcommand{\ppbar}{{p{\bar p}}}
\newcommand{\qqgam}{{q{\bar q}(\gamma)}}
\newcommand{\ff}{{f{\bar f}}}
\newcommand{\lplm}{{\elll^{+}\elll^{-}}}
\newcommand{\bb}{{b{\bar b}}}
\newcommand{\cc}{{c{\bar c}}}
\newcommand{\ee}{{e^{+}e^{-}}}
\newcommand{\mumu}{{\mu^{+}\mu^{-}}}
\newcommand{\tautau}{{\tau^{+}\tau^{-}}}
\newcommand{\eemm}{{\ee \ra \mumu}}
\newcommand{\eemmg}{{\ee \ra \mumu \gamma}}
\newcommand{\eett}{{\ee \ra \tautau}}
\newcommand{\eeee}{{\ee \ra \ee}}
\newcommand{\eell}{{\ee \ra \lplm}}
\newcommand{\nl}{{\nu_{\ell}}}
\newcommand{\nbl}{{{\overline{\nu}}_{\ell}}}
\newcommand{\nmu}{{\nu_{\mu}}}
\newcommand{\nbmu}{{{\overline{\nu}}_{\mu}}}
\newcommand{\nel}{{\nu_{e}}}
\newcommand{\nbel}{{{\overline{\nu}}_{e}}}
\newcommand{\ntau}{{\nu_{\tau}}}
\newcommand{\nbtau}{{{\overline{\nu}}_{\tau}}}
\newcommand{\ppos}{{\pi^{+}}}
\newcommand{\pneg}{{\pi^{-}}}
\newcommand{\pzer}{{\pi^{0}}}
\newcommand{\ppm}{{\pi^{\pm }}}
\newcommand{\Zo}{{\mathrm{Z}^{0}}}
\newcommand{\Zostar}{{\mathrm{Z}^{0\ast}}}
\newcommand{\Wpm}{{\mathrm{W}^\pm}}
\newcommand{\Wp}{{\mathrm{W}^{+}}}
\newcommand{\Wm}{{\mathrm{W}^{-}}}
\newcommand{\W}{{\mathrm{W}}}
\newcommand{\Z}{{\mathrm{Z}}}
\newcommand{\ZZ}{{\mathrm{ZZ}}}
\newcommand{\Zee}{{\mathrm{Z \ee}}}
\newcommand{\Wen}{{\mathrm{W e \nel}}}
\newcommand{\Zgam}{{\mathrm{Z/\gamma}}}
\newcommand{\Ho}{{\mathrm{H}}}
\newcommand{\WW}{{\mathrm{ W^{+} W^{-} }}}

\newcommand{\qqb}{{q{\bar q'}}}
\newcommand{\qbq}{{{\bar q}q'}}
\newcommand{\len}{{\ell \nbl}}
\newcommand{\mn}{{\mu \nbmu}}
\newcommand{\en}{{e \nbel}}
\newcommand{\tn}{{\tau \nbtau}}
\newcommand{\nmn}{{{\overline{\mu}} \nmu}}
\newcommand{\nen}{{{\overline{e}} \nel}}
\newcommand{\ntn}{{{\overline{\tau}} \ntau}}

\newcommand{\qqqq}{{\mathrm{ \qqb \qbq}}} 
\newcommand{\lnqq}{{\mathrm{ \len  \qqb}}}
\newcommand{\enqq}{{\mathrm{ \en  \qqb}}}
\newcommand{\mnqq}{{\mathrm{ \mn  \qqb}}}
\newcommand{\tnqq}{{\mathrm{ \tn  \qqb}}}
\newcommand{\lnln}{{\ell \nbl {\overline{\ell}} \nl}}
\newcommand{\enen}{{\en \nen }}  
\newcommand{\mnmn}{{\mn \nmn}}
\newcommand{\tntn}{{\tn \ntn}}
\newcommand{\enmn}{{\en \nmn}}
\newcommand{\entn}{{\en \ntn}}
\newcommand{\mntn}{{\mn \ntn}}

\newcommand{\mw}{{\mathrm{M_W}}}
\newcommand{\mwone}{{\mathrm{m_{\W 1}}}}
\newcommand{\mwtwo}{{\mathrm{m_{\W 2}}}}
\newcommand{\smw}{{\mathrm{m_W}}}
\newcommand{\mwav}{{\mathrm{\bar{m}_W}}}

\newcommand{\mz}{{\mathrm{M_Z}}}
\newcommand{\MZrp}[1]{{\mathrm{M_{\mbox{\tiny{Z}}}\!\!^{#1}}}}
\newcommand{\gz}{{\Gamma_{\mathrm{Z}}}}
\newcommand{\tz}{{\tau_{\mathrm{Z}}}}

\newcommand{\gw}{{\Gamma_{\mathrm{W}}}}
\newcommand{\mh}{{\mathrm{M_H}}}
\newcommand{\mt}{{\mathrm{m_t}}}
\newcommand{\ssqtw}{{\sin^{2}\!\theta_{\mathrm{W}}}}
\newcommand{\csqtw}{{\cos^{2}\!\theta_{\mathrm{W}}}}
\newcommand{\stw}{{\sin\theta_{\mathrm{W}}}}
\newcommand{\ctw}{{\cos\theta_{\mathrm{W}}}}
\newcommand{\ssqtwef}{{\sin}^{2}\theta_{\mathrm{W}}^{\mathrm{eff}}}
\newcommand{\ssqtfef}{{{\sin}^{2}\theta^{\mathrm{eff}}_{f}}}
\newcommand{\ssqtlef}{{{\sin}^{2}\theta^{\mathrm{eff}}_{l}}}
\newcommand{\csqtwef}{{{\cos}^{2}\theta_{\mathrm{W}}^{\mathrm{eff}}}}
\newcommand{\stwef}{\sin\theta_{\mathrm{W}}^{\mathrm{eff}}}
\newcommand{\ctwef}{\cos\theta_{\mathrm{W}}^{\mathrm{eff}}}
\newcommand{\gv}{{g_{\mbox{\tiny V}}}}
\newcommand{\ga}{{g_{\mbox{\tiny A}}}}
\newcommand{\gvel}{{g_{\mbox{\tiny V}}^{e}}}
\newcommand{\gael}{{g_{\mbox{\tiny A}}^{e}}}
\newcommand{\gvmu}{{g_{\mbox{\tiny V}}^{\mu}}}
\newcommand{\gamu}{{g_{\mbox{\tiny A}}^{\mu}}}
\newcommand{\gvf}{{g_{\mbox{\tiny V}}^{f}}}
\newcommand{\gaf}{{g_{\mbox{\tiny A}}^{f}}}
\newcommand{\gvl}{{g_{\mbox{\tiny V}}^{l}}}
\newcommand{\gal}{{g_{\mbox{\tiny A}}^{l}}}
\newcommand{\ghvf}{{\hat{g}_{\mbox{\tiny V}}^{f}}}
\newcommand{\ghaf}{{\hat{g}_{\mbox{\tiny A}}^{f}}}
\newcommand{\gvh}{{\hat{g}_{\mbox{\tiny V}}}}
\newcommand{\gah}{{\hat{g}_{\mbox{\tiny A}}}}
\newcommand{\thw}{{\theta_{\mbox{\mathrm{W}}}}}
\newcommand{\GF}{{G_{\mbox{\tiny F}}}}
\newcommand{\Vub}{{\mathrm{V}_{ub}}}
\newcommand{\Vcb}{{\mathrm{V}_{cb}=}}

\newcommand{\mgen}{{\mathrm{M_{gen}}}}
\newcommand{\ordalph}{{\mathcal{O}(\alpha)}}
\newcommand{\ordalsq}{{\mathcal{O}(\alpha^{2})}}
\newcommand{\ordalcb}{{\mathcal{O}(\alpha^{3})}}
\newcommand{\sigtot}{{\sigma_{\mbox{\tiny TOT}}}}
\newcommand{\sigf}{{\sigma_{\mbox{\tiny F}}}}
\newcommand{\sigb}{{\sigma_{\mbox{\tiny B}}}}
\newcommand{\dsigf}{{\delta\sigma_{\mbox{\tiny F}}}}
\newcommand{\dsigb}{{\delta\sigma_{\mbox{\tiny B}}}}
\newcommand{\dsfbi}{{\delta\sigma_{\mbox{\tiny fb}}^{\mbox{\tiny int}}}}
\newcommand{\AFB}{{A_{\mbox{\tiny FB}}}}
\newcommand{\Afbmm}{{A_{\mbox{\tiny FB}}^{\mbox{\tiny $\mu\mu$}}}}
\newcommand{\Afbtt}{{A_{\mbox{\tiny FB}}^{\mbox{\tiny $\tau\tau$}}}}
\newcommand{\APOL}{{A_{\mbox{\tiny POL}}}}
\newcommand{\pwff}{{\Gamma_{ff}}}
\newcommand{\pwee}{{\Gamma_{ee}}}
\newcommand{\pwmm}{{\Gamma_{\mu\mu}}}
\newcommand{\Af}{{\mathcal{A}_{f}}}
\newcommand{\Ael}{{\mathcal{A}_{e}}}
\newcommand{\Amu}{{\mathcal{A}_{\mu}}}
\newcommand{\dafbint}{{\delta A_{\mbox{\tiny FB}}^{\mbox{\tiny int}}}}
\newcommand{\vcs}{{\left| V_{cs} \right|} }

\newcommand{\Ephot}{{E_{\gamma}}}
\newcommand{\Ebeam}{{E_{\mbox{\tiny BEAM}}}}
\newcommand{\sqs}{{\protect\sqrt{s}}}
\newcommand{\sprime}{{\protect\sqrt{s^{\prime}}}}
\newcommand{\pT}{{\mathrm{p_T}}}
\newcommand{\mmu}{{m_{\mu}}}
\newcommand{\mb}{{m_{b}}}
\newcommand{\thacop}{{\theta_{\mbox{\tiny acop}}}}
\newcommand{\thacol}{{\theta_{\mbox{\tiny acol}}}}
\newcommand{\prad}{{p_{\mbox{\tiny rad}}}}
\newcommand{\lambdabar}{{\lambda \! \! \! \! {\raisebox{+.5ex}{$-$}}}}

\newcommand{\rprg}{{r_{\mbox{\tiny p}}}}
\newcommand{\zprg}{{z_{\mbox{\tiny p}}}}
\newcommand{\thprg}{{\theta_{\mbox{\tiny p}}}}
\newcommand{\phprg}{{\phi_{\mbox{\tiny p}}}}

\newcommand{\mrad}{{\mathrm{mrad}}}
\newcommand{\rad}{{\mathrm{rad}}}
\newcommand{\dgr}{{^\circ}}
\newcommand{\TeV}{{\mathrm{TeV}}}
\newcommand{\GeV}{{\mathrm{GeV}}}
\newcommand{\GeVm}{{\mathrm{GeV/c{^2}}}}
\newcommand{\GeVp}{{\mathrm{GeV/c}}}
\newcommand{\MeV}{{\mathrm{MeV}}}
\newcommand{\MeVp}{{\mathrm{MeV/c}}}
\newcommand{\MeVm}{{\mathrm{MeV/c{^2}}}}
\newcommand{\KeV}{{\mathrm{KeV}}}
\newcommand{\eV}{{\mathrm{eV}}}
\newcommand{\um}{{\mathrm{m}}}
\newcommand{\umm}{{\mathrm{mm}}}
\newcommand{\uum}{{\mu{\mathrm m}}}
\newcommand{\ucm}{{\mathrm{cm}}}
\newcommand{\ufm}{{\mathrm{fm}}}
\newcommand{\umicrom}{{\mathrm{\mu m}}}
\newcommand{\us}{{\mathrm{s}}}
\newcommand{\ums}{{\mathrm{ms}}}
\newcommand{\uus}{{\mu{\mathrm{s}}}}
\newcommand{\uns}{{\mathrm{ns}}}
\newcommand{\ups}{{\mathrm{ps}}}
\newcommand{\uub}{{\mu{\mathrm{b}}}}
\newcommand{\unb}{{\mathrm{nb}}}
\newcommand{\upb}{{\mathrm{pb}}}
\newcommand{\ipb}{{\mathrm{pb^{-1}}}}
\newcommand{\ifb}{{\mathrm{fb^{-1}}}}
\newcommand{\inb}{{\mathrm{nb^{-1}}}}

\newcommand{\eg}{\mbox{\itshape e.g.}}
\newcommand{\ie}{\mbox{\itshape i.e.}}
\newcommand{\etal}{{\slshape et al\/}\ }
\newcommand{\etc}{\mbox{\itshape etc}}
\newcommand{\cf}{\mbox{\itshape cf.}}
\newcommand{\ffp}{\mbox{\itshape ff}}
\newcommand{\cm}{\mbox{c.m.}}
\newcommand{\ALEPH}{\mbox{\scshape Aleph}}
\newcommand{\DELPHI}{\mbox{\scshape Delphi}}
\newcommand{\OPAL}{\mbox{\scshape Opal}}
\newcommand{\LTHREE}{\mbox{\scshape L3}}
\newcommand{\CERN}{\mbox{\scshape Cern}}
\newcommand{\LEP}{\mbox{\scshape Lep}}
\newcommand{\LEPONE}{\mbox{\scshape Lep1}}
\newcommand{\LEPTWO}{\mbox{\scshape Lep2}}
\newcommand{\CDF}{\mbox{\scshape Cdf}}
\newcommand{\DO}{\mbox{\scshape D0}}
\newcommand{\SLD}{\mbox{\scshape Sld}}
\newcommand{\CLEO}{\mbox{\scshape Cleo}}
\newcommand{\UAONE}{\mbox{\scshape Ua1}}
\newcommand{\UATWO}{\mbox{\scshape Ua2}}
\newcommand{\TEVATRON}{\mbox{\scshape Tevatron}}
\newcommand{\LHC}{\mbox{\scshape LHC}}
\newcommand{\KORALZ}{\mbox{\ttfamily KORALZ}}
\newcommand{\KORALW}{\mbox{\ttfamily KORALW}}
\newcommand{\ZFITTER}{\mbox{\ttfamily ZFITTER}}
\newcommand{\GENTLE}{\mbox{\ttfamily GENTLE}}
\newcommand{\DELANA}{\mbox{\ttfamily DELANA}}
\newcommand{\DELSIM}{\mbox{\ttfamily DELSIM}}
\newcommand{\DYMU}{\mbox{\ttfamily DYMU3}}
\newcommand{\TANAGRA}{\mbox{\ttfamily TANAGRA}}
\newcommand{\ZEBRA}{\mbox{\ttfamily ZEBRA}}
\newcommand{\PAW}{\mbox{\ttfamily PAW}}
\newcommand{\WWANA}{\mbox{\ttfamily WWANA}}
\newcommand{\FASTSIM}{\mbox{\ttfamily FASTSIM}}
\newcommand{\PYTHIA}{\mbox{\ttfamily PYTHIA}}
\newcommand{\JETSET}{\mbox{\ttfamily JETSET}}
\newcommand{\TWOGAM}{\mbox{\ttfamily TWOGAM}}
\newcommand{\LUBOEI}{\mbox{\ttfamily LUBOEI}}
\newcommand{\SKI}{\mbox{\ttfamily SK-I}}
\newcommand{\SKII}{\mbox{\ttfamily SK-II}}
\newcommand{\ARIADNE}{\mbox{\ttfamily ARIADNE}}
\newcommand{\ARII}{\mbox{\ttfamily AR-II}}
\newcommand{\VNI}{\mbox{\ttfamily VNI}}
\newcommand{\HERWIG}{\mbox{\ttfamily HERWIG}}
\newcommand{\EXCALIBUR}{\mbox{\ttfamily EXCALIBUR}}
\newcommand{\WPHACT}{\mbox{\ttfamily WPHACT}}
\newcommand{\RACOONWW}{\mbox{\ttfamily RACOONWW}}
\newcommand{\YFSWW}{\mbox{\ttfamily YFSWW}}
\newcommand{\QEDPS}{\mbox{\ttfamily QEDPS}}
\newcommand{\CCTHREE}{\mbox{\ttfamily CC03}}
\newcommand{\NCEIGHT}{\mbox{\ttfamily NC08}}
\newcommand{\LUCLUS}{\mbox{\ttfamily LUCLUS}}
\newcommand{\JADE}{\mbox{\ttfamily JADE}}
\newcommand{\DURHAM}{\mbox{\ttfamily DURHAM}}
\newcommand{\CAMJET}{\mbox{\ttfamily CAMJET}}
\newcommand{\DICLUS}{\mbox{\ttfamily DICLUS}}
\newcommand{\CONE}{\mbox{\ttfamily CONE}}
\newcommand{\PUFITC}{\mbox{\ttfamily PUFITC+}}
\newcommand{\PHDST}{\mbox{\ttfamily PHDST}}
\newcommand{\SKELANA}{\mbox{\ttfamily SKELANA}}
\newcommand{\DAFNE}{\mbox{\ttfamily DAFNE}}
\newcommand{\CERNLIB}{\mbox{\ttfamily CERNLIB}}
\newcommand{\MINUIT}{\mbox{\ttfamily MINUIT}}
\newcommand{\REMCLU}{\mbox{\ttfamily REMCLU}}
\newcommand{\DST}{\mbox{\ttfamily DST}}
\newcommand{\XSDST}{\mbox{\ttfamily XShortDST}}
\newcommand{\FDST}{\mbox{\ttfamily FullDST}}

\newcommand{\spot}{\mbox{$\bullet \;$}}

\newcommand{\eps}{{\epsilon}}
\newcommand{\erreps}{{\sigma_{\epsilon}}}
\newcommand{\errepsp}{{\sigma_{\epsilon}^{+}}}
\newcommand{\errepsm}{{\sigma_{\epsilon}^{-}}}
\newcommand{\Lmb}{{\Lambda}}
\newcommand{\lmb}{{\lambda}}
\newcommand{\Lmbsq}{{\Lambda^{2}}}
\newcommand{\Lmbisq}{{1/\Lambda^{2}}}
\newcommand{\Lmbpm}{{\Lambda^{\pm}}}
\newcommand{\Lmbp}{{\Lambda^{+}}}
\newcommand{\Lmbm}{{\Lambda^{-}}}
\newcommand{\gc}{{g_{c}}}
\newcommand{\etaij}{{\eta_{ij}}}
\newcommand{\IJ}{{\mathrm{IJ}}}
\newcommand{\IJpm}[1]{{\mathrm{IJ}^{(#1)}}}

\newcommand{\gev}{{\mathrm{GeV}}}
\newcommand{\wpm}{{\mathrm{W}^\pm}}
\newcommand{\wm}{{\mathrm{W}^{-}}}
\newcommand{\w}{{\mathrm{W}}}
\newcommand{\z}{{\mathrm{Z}}}
\newcommand{\ww}{{\mathrm{ W^{+} W^{-} }}}
\newcommand{\tev}{{\mathrm{TeV}}}
\newcommand{\lep}{\mbox{\scshape Lep}}
\newcommand{\lepone}{\mbox{\scshape Lep1}}
\newcommand{\leptwo}{\mbox{\scshape Lep2}}
\newcommand{\delphi}{\mbox{\scshape Delphi}}
\newcommand{\excalibur}{\mbox{\ttfamily EXCALIBUR}}
\newcommand{\wwana}{\mbox{\ttfamily WWANA}}
\newcommand{\fastsim}{\mbox{\ttfamily FASTSIM}}
\newcommand{\phdst}{\mbox{\ttfamily PHDST}}
\newcommand{\skelana}{\mbox{\ttfamily SKELANA}}
\newcommand{\dafne}{\mbox{\ttfamily DAFNE}}
\newcommand{\delana}{\mbox{\ttfamily DELANA}}
\newcommand{\zebra}{\mbox{\ttfamily ZEBRA}}
\newcommand{\delsim}{\mbox{\ttfamily DELSIM}}
\newcommand{\pythia}{\mbox{\ttfamily PYTHIA}}
\newcommand{\jetset}{\mbox{\ttfamily JETSET}}
\newcommand{\ariadne}{\mbox{\ttfamily ARIADNE}}
\newcommand{\cdf}{\mbox{\scshape CDF}}
\newcommand{\gentle}{\mbox{\ttfamily GENTLE}}

A measurement of the W mass and width has been performed by the \DELPHI\
collaboration using the data collected during 1998. The data sample has an
integrated luminosity of 155 $\ipb$ and an average centre-of-mass energy of
188.6 $\GeV$ .
Results are obtained by applying the method of direct reconstruction of the mass of the W from its decay products in both the $\WW\ \rightarrow\ \lnqq$ and $\WW\ \rightarrow\ \qqqq$ channels. The W mass result for the 1998 data set is
\begin{center}  $\mw =  80.387 \pm 0.087 (stat) \pm 0.034 (syst) \pm 0.017 (LEP) \pm 0.035 (FSI) ~ \GeVm$,  \end{center}

where $FSI$ represents the uncertainty due to final state interaction effects in the $\qqqq$ channel, and $LEP$ represents that arising from the knowledge of the beam energy of the accelerator. Combining this result with those previously published by the \DELPHI\ collaboration gives the result
\begin{center} $\mw  =  80.359 \pm 0.074 (stat) \pm 0.032 (syst) \pm 0.017 (LEP) \pm 0.033 (FSI) ~ \GeVm$. \end{center}
The combined value for the W width is
\begin{center} $\gw  =  2.266 \pm 0.176 (stat) \pm 0.056 (syst) \pm 0.052 (FSI) ~ \GeVm$. \end{center}


\end{shortabs}
\vfill
\begin{center}
\DpSubmit \ \\ 
\DpComment \ \\
\DpEMail \ \\
\end{center}
\vfill
\clearpage
\headsep 10.0pt
\addtolength{\textheight}{10mm}
\addtolength{\footskip}{-5mm}
\begingroup
%
\newcommand{\DpName}[2]{\hbox{#1$^{\ref{#2}}$},\hfill}
\newcommand{\DpNameTwo}[3]{\hbox{#1$^{\ref{#2},\ref{#3}}$},\hfill}
\newcommand{\DpNameThree}[4]{\hbox{#1$^{\ref{#2},\ref{#3},\ref{#4}}$},\hfill}
\newskip\Bigfill \Bigfill = 0pt plus 1000fill
\newcommand{\DpNameLast}[2]{\hbox{#1$^{\ref{#2}}$}\hspace{\Bigfill}}
%
\footnotesize
\noindent
\DpName{P.Abreu}{LIP}
\DpName{W.Adam}{VIENNA}
\DpName{T.Adye}{RAL}
\DpName{P.Adzic}{DEMOKRITOS}
\DpName{Z.Albrecht}{KARLSRUHE}
\DpName{T.Alderweireld}{AIM}
\DpName{G.D.Alekseev}{JINR}
\DpName{R.Alemany}{CERN}
\DpName{T.Allmendinger}{KARLSRUHE}
\DpName{P.P.Allport}{LIVERPOOL}
% ...
\normalsize
\endgroup
\titlefoot{Department of Physics and Astronomy, Iowa State
     University, Ames IA 50011-3160, USA
    \label{AMES}}
\titlefoot{Physics Department, Univ. Instelling Antwerpen,
     Universiteitsplein 1, B-2610 Antwerpen, Belgium \\
     \indent~~and IIHE, ULB-VUB,
     Pleinlaan 2, B-1050 Brussels, Belgium \\
     \indent~~and Facult\'e des Sciences,
     Univ. de l'Etat Mons, Av. Maistriau 19, B-7000 Mons, Belgium
    \label{AIM}}
\titlefoot{Physics Laboratory, University of Athens, Solonos Str.
     104, GR-10680 Athens, Greece
    \label{ATHENS}}
\titlefoot{Department of Physics, University of Bergen,
     All\'egaten 55, NO-5007 Bergen, Norway
    \label{BERGEN}}
\titlefoot{Dipartimento di Fisica, Universit\`a di Bologna and INFN,
     Via Irnerio 46, IT-40126 Bologna, Italy
    \label{BOLOGNA}}
\titlefoot{Centro Brasileiro de Pesquisas F\'{\i}sicas, rua Xavier Sigaud 150,
     BR-22290 Rio de Janeiro, Brazil \\
     \indent~~and Depto. de F\'{\i}sica, Pont. Univ. Cat\'olica,
     C.P. 38071 BR-22453 Rio de Janeiro, Brazil \\
     \indent~~and Inst. de F\'{\i}sica, Univ. Estadual do Rio de Janeiro,
     rua S\~{a}o Francisco Xavier 524, Rio de Janeiro, Brazil
    \label{BRASIL}}
\titlefoot{Comenius University, Faculty of Mathematics and Physics,
     Mlynska Dolina, SK-84215 Bratislava, Slovakia
    \label{BRATISLAVA}}
\titlefoot{Coll\`ege de France, Lab. de Physique Corpusculaire, IN2P3-CNRS,
     FR-75231 Paris Cedex 05, France
    \label{CDF}}
\titlefoot{CERN, CH-1211 Geneva 23, Switzerland
    \label{CERN}}
\titlefoot{Institut de Recherches Subatomiques, IN2P3 - CNRS/ULP - BP20,
     FR-67037 Strasbourg Cedex, France
    \label{CRN}}
\titlefoot{Now at DESY-Zeuthen, Platanenallee 6, D-15735 Zeuthen, Germany
    \label{DESY}}
\titlefoot{Institute of Nuclear Physics, N.C.S.R. Demokritos,
     P.O. Box 60228, GR-15310 Athens, Greece
    \label{DEMOKRITOS}}
\titlefoot{FZU, Inst. of Phys. of the C.A.S. High Energy Physics Division,
     Na Slovance 2, CZ-180 40, Praha 8, Czech Republic
    \label{FZU}}
\titlefoot{Currently at DPNC, University of Geneva,
     Quai Ernest-Ansermet 24, CH-1211, Geneva, Switzerland
    \label{GENEVA}}
\titlefoot{Dipartimento di Fisica, Universit\`a di Genova and INFN,
     Via Dodecaneso 33, IT-16146 Genova, Italy
    \label{GENOVA}}
\titlefoot{Institut des Sciences Nucl\'eaires, IN2P3-CNRS, Universit\'e
     de Grenoble 1, FR-38026 Grenoble Cedex, France
    \label{GRENOBLE}}
\titlefoot{Helsinki Institute of Physics, HIP,
     P.O. Box 9, FI-00014 Helsinki, Finland
    \label{HELSINKI}}
\titlefoot{Joint Institute for Nuclear Research, Dubna, Head Post
     Office, P.O. Box 79, RU-101 000 Moscow, Russian Federation
    \label{JINR}}
\titlefoot{Institut f\"ur Experimentelle Kernphysik,
     Universit\"at Karlsruhe, Postfach 6980, DE-76128 Karlsruhe,
     Germany
    \label{KARLSRUHE}}
\titlefoot{Institute of Nuclear Physics and University of Mining and Metalurgy,
     Ul. Kawiory 26a, PL-30055 Krakow, Poland
    \label{KRAKOW}}
\titlefoot{Universit\'e de Paris-Sud, Lab. de l'Acc\'el\'erateur
     Lin\'eaire, IN2P3-CNRS, B\^{a}t. 200, FR-91405 Orsay Cedex, France
    \label{LAL}}
\titlefoot{LIP, IST, FCUL - Av. Elias Garcia, 14-$1^{o}$,
     PT-1000 Lisboa Codex, Portugal
    \label{LIP}}
\titlefoot{Department of Physics, University of Liverpool, P.O.
     Box 147, Liverpool L69 3BX, UK
    \label{LIVERPOOL}}
\titlefoot{LPNHE, IN2P3-CNRS, Univ.~Paris VI et VII, Tour 33 (RdC),
     4 place Jussieu, FR-75252 Paris Cedex 05, France
    \label{LPNHE}}
\titlefoot{Department of Physics, University of Lund,
     S\"olvegatan 14, SE-223 63 Lund, Sweden
    \label{LUND}}
\titlefoot{Universit\'e Claude Bernard de Lyon, IPNL, IN2P3-CNRS,
     FR-69622 Villeurbanne Cedex, France
    \label{LYON}}
\titlefoot{Univ. d'Aix - Marseille II - CPP, IN2P3-CNRS,
     FR-13288 Marseille Cedex 09, France
    \label{MARSEILLE}}
\titlefoot{Dipartimento di Fisica, Universit\`a di Milano and INFN-MILANO,
     Via Celoria 16, IT-20133 Milan, Italy
    \label{MILANO}}
\titlefoot{Dipartimento di Fisica, Univ. di Milano-Bicocca and
     INFN-MILANO, Piazza delle Scienze 2, IT-20126 Milan, Italy
    \label{MILANO2}}
\titlefoot{IPNP of MFF, Charles Univ., Areal MFF,
     V Holesovickach 2, CZ-180 00, Praha 8, Czech Republic
    \label{NC}}
\titlefoot{NIKHEF, Postbus 41882, NL-1009 DB
     Amsterdam, The Netherlands
    \label{NIKHEF}}
\titlefoot{National Technical University, Physics Department,
     Zografou Campus, GR-15773 Athens, Greece
    \label{NTU-ATHENS}}
\titlefoot{Physics Department, University of Oslo, Blindern,
     NO-1000 Oslo 3, Norway
    \label{OSLO}}
\titlefoot{Dpto. Fisica, Univ. Oviedo, Avda. Calvo Sotelo
     s/n, ES-33007 Oviedo, Spain
    \label{OVIEDO}}
\titlefoot{Department of Physics, University of Oxford,
     Keble Road, Oxford OX1 3RH, UK
    \label{OXFORD}}
\titlefoot{Dipartimento di Fisica, Universit\`a di Padova and
     INFN, Via Marzolo 8, IT-35131 Padua, Italy
    \label{PADOVA}}
\titlefoot{Rutherford Appleton Laboratory, Chilton, Didcot
     OX11 OQX, UK
    \label{RAL}}
\titlefoot{Dipartimento di Fisica, Universit\`a di Roma II and
     INFN, Tor Vergata, IT-00173 Rome, Italy
    \label{ROMA2}}
\titlefoot{Dipartimento di Fisica, Universit\`a di Roma III and
     INFN, Via della Vasca Navale 84, IT-00146 Rome, Italy
    \label{ROMA3}}
\titlefoot{DAPNIA/Service de Physique des Particules,
     CEA-Saclay, FR-91191 Gif-sur-Yvette Cedex, France
    \label{SACLAY}}
\titlefoot{Instituto de Fisica de Cantabria (CSIC-UC), Avda.
     los Castros s/n, ES-39006 Santander, Spain
    \label{SANTANDER}}
\titlefoot{Inst. for High Energy Physics, Serpukov
     P.O. Box 35, Protvino, (Moscow Region), Russian Federation
    \label{SERPUKHOV}}
\titlefoot{J. Stefan Institute, Jamova 39, SI-1000 Ljubljana, Slovenia
     and Laboratory for Astroparticle Physics,\\
     \indent~~Nova Gorica Polytechnic, Kostanjeviska 16a, SI-5000 Nova Gorica, Slovenia, \\
     \indent~~and Department of Physics, University of Ljubljana,
     SI-1000 Ljubljana, Slovenia
    \label{SLOVENIJA}}
\titlefoot{Fysikum, Stockholm University,
     Box 6730, SE-113 85 Stockholm, Sweden
    \label{STOCKHOLM}}
\titlefoot{Dipartimento di Fisica Sperimentale, Universit\`a di
     Torino and INFN, Via P. Giuria 1, IT-10125 Turin, Italy
    \label{TORINO}}
\titlefoot{Dipartimento di Fisica, Universit\`a di Trieste and
     INFN, Via A. Valerio 2, IT-34127 Trieste, Italy \\
     \indent~~and Istituto di Fisica, Universit\`a di Udine,
     IT-33100 Udine, Italy
    \label{TU}}
\titlefoot{Univ. Federal do Rio de Janeiro, C.P. 68528
     Cidade Univ., Ilha do Fund\~ao
     BR-21945-970 Rio de Janeiro, Brazil
    \label{UFRJ}}
\titlefoot{Department of Radiation Sciences, University of
     Uppsala, P.O. Box 535, SE-751 21 Uppsala, Sweden
    \label{UPPSALA}}
\titlefoot{IFIC, Valencia-CSIC, and D.F.A.M.N., U. de Valencia,
     Avda. Dr. Moliner 50, ES-46100 Burjassot (Valencia), Spain
    \label{VALENCIA}}
\titlefoot{Institut f\"ur Hochenergiephysik, \"Osterr. Akad.
     d. Wissensch., Nikolsdorfergasse 18, AT-1050 Vienna, Austria
    \label{VIENNA}}
\titlefoot{Inst. Nuclear Studies and University of Warsaw, Ul.
     Hoza 69, PL-00681 Warsaw, Poland
    \label{WARSZAWA}}
\titlefoot{Fachbereich Physik, University of Wuppertal, Postfach
     100 127, DE-42097 Wuppertal, Germany
    \label{WUPPERTAL}}
\addtolength{\textheight}{-10mm}
\addtolength{\footskip}{5mm}
\clearpage
\headsep 30.0pt
\end{titlepage}
%
\pagenumbering{arabic} 
\setcounter{footnote}{0} %
\large
%
%
%


\def\Journal#1#2#3#4{{#1}{\bf #2}(#4) #3}
\def\PLB{{ Phys. Lett.}  \bf B}
\def\EUR{{ Eur. Phys. J.} \bf C}
\def\PRL{{ Phys. Rev. Lett.} }
\def\NIMA{{Nucl. Instr. and Meth.} \bf A}
\def\PRD{{ Phys. Rev.} \bf D}
\def\ZPC{{ Zeit. Phys.} \bf C}
\def\CPC{{Comput. Phys. Commun.} }

\section{Introduction}
The W mass has been measured by the \DELPHI\ collaboration using the data collected during 1998. This data sample has allowed a significant improvement in the accuracy of the collaboration's W mass determination as the integrated luminosity is more than twice that on which previous \DELPHI\ results are based \cite{delpaper161,delpaper172,delpaper183}. The W mass has also been determined by the other \LEP\ collaborations \cite{lepmw} and at hadron colliders \cite{hadmw}. Using the same reconstruction method as for the W mass, results on the direct measurement of the W width are also obtained in this paper, and can be compared with those of the other \LEP\ collaborations and of the \CDF\ collaboration \cite{cdfwidth}.  

Section \ref{sec:data} of this paper describes the characteristics of the 1998 data sample and of the event generators used in this analysis.

The analysis is performed through the direct reconstruction of the mass of the W from its decay products in the $\WW \rightarrow \qqqq$ (fully-hadronic) and
 $\WW \rightarrow \lnqq$ (semi-leptonic) decay channels.
The applied methods are described in Sec. \ref{sec:anal}, and have been
refined from those in previous publications. Results are now reported for $\tnqq$ events, and in both the fully-hadronic and semi-leptonic channels improvements in the handling of initial-state radiation (ISR) have been included.

The systematic error evaluation described in Sec. \ref{sec:syst} has increased in sophistication from that previously reported. New techniques have been applied, such as the use of mixed Lorentz boosted Z's, and a wider range of higher precision simulation studies have been performed.  

The results of this analysis are reported in Sec. \ref{sec:res}, and are combined with the previous \DELPHI\ results.


\section{Data and Simulation Samples}
\label{sec:data}

\subsection{Data}

A detailed description of the \DELPHI\ apparatus and its performance can be found in \cite{delphi} \footnote{The co-ordinate system used has the z-axis parallel to the electron beam, and the polar angle calculated with respect to this axis.}. In 1998 the detector was used to record data at the Z peak and at a nominal centre-of-mass energy of 189 $\GeV$.

The Z peak data were recorded before (1.8 $\ipb$) and towards the end (0.6 $\ipb$) of the  high energy data, thus facilitating checks of detector stability. These data were the principal sample used for calibration and alignment of the detector and, in this analysis, assist the study of systematic uncertainties. 

The average centre-of-mass collision energy for the high energy data was $188.6 ~\GeV$.  The luminosity weighted r.m.s. of this value, assessed on a fill by fill basis, was $50 ~\MeV$. In the data sample considered for analysis all the detectors essential for this measurement were required to be fully efficient; the operation of the central tracking detectors was important for all decay channels, in the $\lnqq$ analysis stricter requirements than in the $\qqqq$ channel were placed on the electromagnetic calorimeters. The selected samples correspond to integrated luminosities of $152.9~\ipb$ for the $\lnqq$ analysis and $157.4~\ipb$ for the $\qqqq$ decay channel.

\subsection{Simulation}

The response of the detector to various physical processes was modelled using the simulation program \DELSIM\ \cite{delphi}, which includes modelling of the resolution, granularity and efficiency of the detector components. In addition, detector correction factors, described in Sec. \ref{sec:syst}, were included to improve the description of jets, electrons and muons. For systematic uncertainty studies a fast simulation program, relying on a relatively simple set of smearing and efficiency parametrisations, was also used. 

WW events and all other four-fermion processes were produced using the event generator \EXCALIBUR\ \cite{excal}, with initial-state radiation described using the \QEDPS\ program \cite{qedps}. The W mass ($\mw$) and width ($\gw$) definition used throughout this paper correspond to a W propagator with an $s$-dependent width. The background process $\ee \rightarrow\  \qqgam$ was simulated with the \PYTHIA\ 5.7 \cite{pythia} event generator. Two photon backgrounds were studied using the \TWOGAM\ generator \cite{twogam}. The fragmentation of all events was performed using \JETSET\ 7.4 \cite{pythia} tuned to the \DELPHI\ \LEP1\ data \cite{deltune}. Systematic error checks were performed using other generators and variations in the fragmentation tuning as described in Sec. \ref{sec:syst}. The systematics section also reports \DELPHI\ results for the common \LEP\ event samples produced in the context of the \LEP\ WW Workshops: the event generation was performed by \ALEPH\ using \KORALW\ 1.21 \cite{koralw}, and these events were then passed through the \DELSIM\ program.

\section{Analysis Method}
\label{sec:anal}
\subsection{Semi-Leptonic Decay Channel}

The fitting procedure presented here is a development of that 
described in~\cite{delpaper183} for the $\enqq$ and $\mnqq$ channels, 
where the fitting function now includes a description of the ISR
spectrum of WW events. In addition, we present here an analysis of
 the $\tnqq$ channel and the event selection in all semi-leptonic
 channels is now based on a neural network.

\subsubsection{Event Selection}
 
{\bf{Lepton Identification}}
\vspace{0.4cm}

     Charged particles were identified as muons if they were associated
 with a hit in the muon chambers, or had an energy deposit 
 in the hadron calorimeter that was consistent with a minimum ionising 
 particle. Muon identification was performed in the polar angle range between 10$^{\circ}$ and 170$^{\circ}$.

Electron identification was performed in the polar angle range between
12$^{\circ}$ and 168$^{\circ}$ by selecting charged particles with a characteristic energy deposition in the electromagnetic calorimeters. In the central region of the detector covered by the HPC electromagnetic calorimeter, electrons were selected using the energy over momentum ($E/p$) ratio of the candidate. For lower energy candidates (below 30 $\GeV$) this was supplemented by selection criteria on the shape of the calorimetric shower and a more stringent comparison of the track extrapolation and calorimetric shower positions, while for electrons of higher energy, negligible energy deposition in the hadron calorimeter was required.
In the polar angle region below 36$^{\circ}$ and above 144$^{\circ}$,
 where the momentum resolution is poorer, tracks associated to electromagnetic energy showers above 8 $\GeV$ and negligible hadron calorimeter energy deposits were considered as electrons.

     Tau candidates were obtained by clustering the events into a three-jet 
configuration using the \LUCLUS\ \cite{LUCLUS} algorithm. The jet with the lowest charged
 multiplicity was chosen as the tau candidate. As the tau lepton predominantly decays into a final state with one or three charged particles, only jets containing between one and four charged tracks were selected.     

\vspace{0.2cm}
{\bf{Selection}}
\vspace{0.2cm}

 The event selection was based upon a multi-layer perceptron neural network \cite{neural}. The network was optimised separately for $\enqq$, $\mnqq$, $\tnqq$ candidates containing only one charged particle and $\tnqq$ candidates with several charged particles.
    
     Having removed the lepton candidate in $\enqq$ and $\mnqq$ events, the \LUCLUS\  jet clusterization algorithm (with a $d_{join}$ of $7.5~\GeVp$) was used to cluster the remaining particles. Events containing more than three jets were re-clustered, forcing them into a three-jet configuration. $\tnqq$ events were clustered as the tau candidate and a two-jet system. The events were reconstructed using a constrained fit imposing conservation of four-momentum and equality of the two W masses in the event. As the energy of the tau lepton is unknown, due to the emission of at least one neutrino in its decay, the mass in the $\tnqq$ channel is entirely determined by the jet system.

    The neural network relied upon the characteristic event properties in each decay channel. The input variables included lepton momentum, polar angle of the missing momentum, and the isolation of the lepton candidate from the hadronic system of the event. The electron and muon identification was obtained as a strong or loose tag and the network was optimised separately in each decay channel for these two categories of identification. 

\begin{figure}[ht]
 \begin{center}
 \epsfig{file=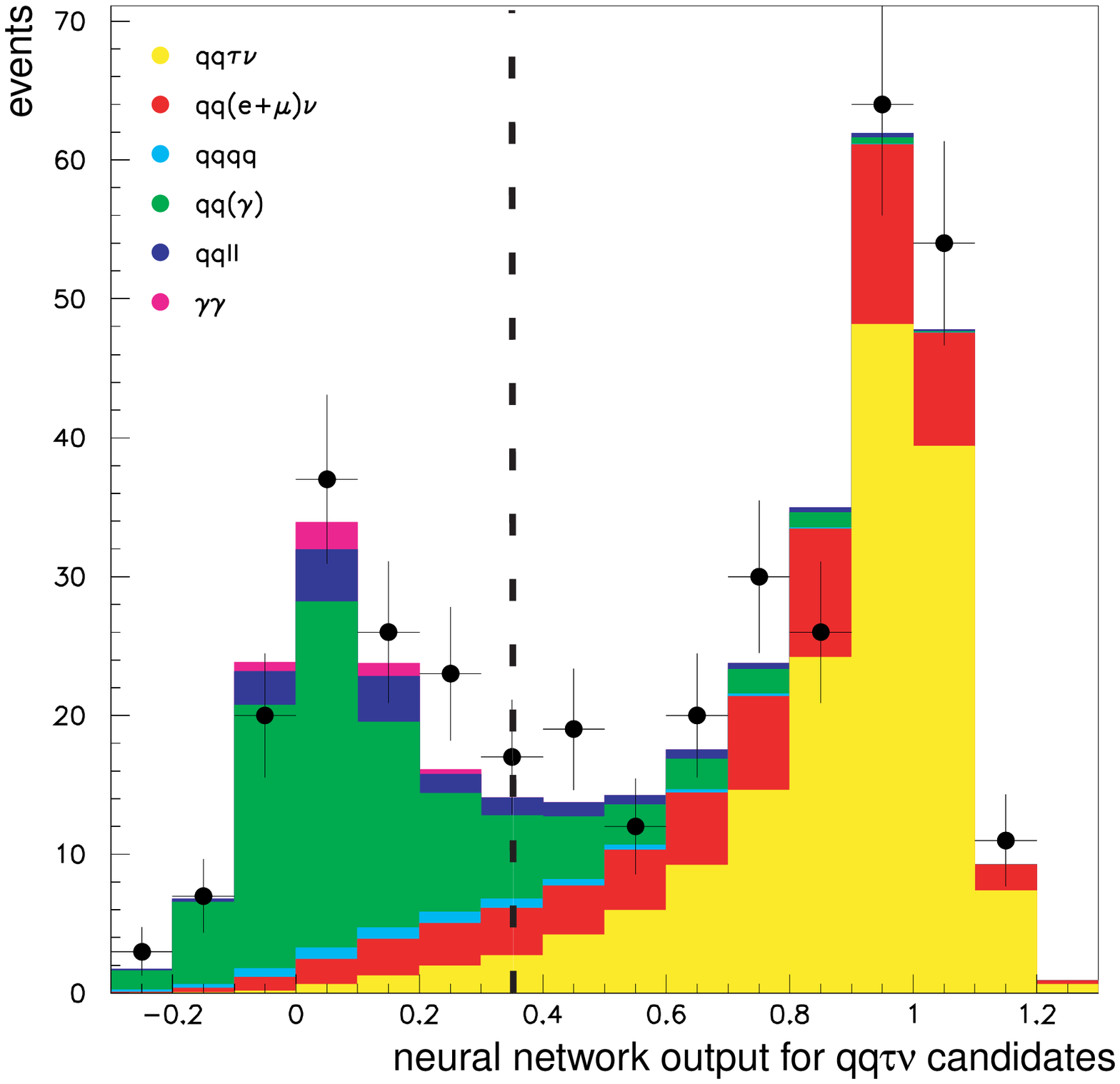,width=0.7\textwidth}
  \caption{The output distribution of the event selection neural network for $\tnqq$ candidates. The shaded areas indicate the contribution from the various simulated states, the data are shown as points with statistical error bars. The value of the selection cut applied is indicated by the dashed line. Note that the order of the simulation contributions in the figure follows that in the key, the $\tnqq$ signal is the first distribution shown and background from two-photon diagrams the last.}

  \label{fig:nntnqq}
 \end{center}
 \end{figure}

The network was tuned on samples of signal and background simulation events, and its performance estimated on independent samples of events. After applying a cut on the network output the selection performance is as shown in Tab. \ref{tab:evtsel}. Events that passed the cut in the muon channel were selected, the remaining events were considered as electron channel candidates and, if they were again rejected, were then analysed under the tau channel hypothesis.
The neural network output in the $\tnqq$ channel is shown in Fig. \ref{fig:nntnqq}.

The tau selection sample contains a significant proportion of other semi-leptonic decays: the composition was estimated from simulation to be $67\%$ $\tnqq$, $19\%$ $\enqq$, $5\%$ $\mnqq$ with the remaining fraction dominated by $q\overline{q}(\gamma)$ background events. This corresponds to a $39\%$ selection efficiency for $\tnqq$ events. Further information on a selection of $\WW$ events in \DELPHI\ with a similar performance is available in \cite{189xsec}.

The fraction of semi-leptonic WW events in the sample was extracted from simulation as a function of the neural network output: this is referred to below as the event purity $P_e$. This feature is particularly useful for the tau selection, where the proportion of background events is highest. 

\subsubsection{Likelihood Function}
 
The following likelihood function was evaluated for each selected events with a reconstructed mass in the range $68-92 ~ \GeVm$ :

\begin{equation}
{\cal L}_e (\mw,\Gamma_{\rm W}) = P_e \cdot S''(m^{fit},\sigma^{fit},\mw,\Gamma_{\rm W}) + (1-P_e)\cdot B(m^{fit}), \end{equation}
 
\noindent where $P_e$ is the event purity, discussed above, $S''$ is the signal function that describes the reconstructed mass distribution of the semi-leptonic W decays, and $B$ is used to describe  background processes. The reconstructed event mass $m^{fit}$ and its estimated error $\sigma^{fit}$ are both obtained from the constrained fit.  The distribution of background events is extracted from simulation as a function of $m^{fit}$.

The signal function $S''$ is defined in terms of $S$,$S'$ as discussed below.
The function $S$ relies on the convolution of three components, using $x$ and $m$ as the dummy integration variables:
    
\begin{equation} S(m^{fit},\sigma^{fit},\mw,\Gamma_{\rm W}) = 
\int_{0}^{\Ebeam} dm~ G [m^{fit}-m,\sigma^{fit}] \int_{0}^{1} dx~ BW_{PS}[m \cdot (1-x),\mw ~ R_{ISR}(x)]. \end{equation}

$BW_{PS}$ is a phase-space corrected relativistic Breit-Wigner distribution (representing the W mass distribution) which is convoluted with the Gaussian function $G$ describing the detector resolution. The width of the Gaussian depends upon the reconstructed mass error obtained in the constrained fit for that event. Details of the $BW_{PS}$ and $G$ terms are given in \cite{delpaper172}. A recent addition to the analysis is the description of the ISR spectrum, which is parametrised as



\begin{equation} R_{ISR}(x_{\gamma}) = \beta x_{\gamma}^{(\beta-1)}, \end{equation}

\noindent where $x_{\gamma}$ is the ratio of the photon energy to the centre-of-mass energy and $\beta$ is calculated from the electromagnetic constant ($\alpha$), the centre-of-mass energy squared ($s$) and the electron mass ($m_{e}$):

\begin{equation} \beta = \frac{2\alpha}{\pi}[\log(s/m_{e}^2)-1]. \end{equation}

Including this ISR term decreases the bias on the fitted W mass by approximately $400~ \MeV$ and improves the expected error by $2 \pm 1 \%$.

   The event selection contains a significant fraction of $\tnqq$ events in
 the electron and muon channel samples, and of $\enqq$ events in
 the tau sample (see Tab. \ref{tab:evtsel}). In the tau channel the mass of the event is determined from the jet system. The behaviour of true $\tnqq$ and $\enqq$ events in this fit are found to be similar. However, in the electron and muon channel samples the behaviour of the $\tnqq$ events is somewhat different to that of the $\enqq$, $\mnqq$ events. The $\tnqq$ events have a worse mass resolution and a small negative bias on the mass. The fraction of tau events, which have been wrongly classified and are contained in the electron and muon channel samples, has been parametrised in bins of the lepton energy and the measured missing mass.  This event impurity $P\tau_{e}$ was then taken into account in the likelihood function for the electron and muon samples, by defining the signal function $S''$ as

 \begin{equation} S'' = (1-P\tau_{e}) \cdot S + P\tau_{e} \cdot S' ,\end{equation}

\noindent where $S'$ is analogous to $S$, but with the width of the Gaussian resolution  function increased according to simulation studies. All remaining biases in the analysis due to using this approximate likelihood description are corrected for in the calibration procedure as described in Sec. \ref{sec:massextrac}.

\subsection{Fully-Hadronic Decay Channel}

The analysis of the fully-hadronic channel was based on that applied in \cite{delpaper183}. However, the implementation now relies on kinematic fits with four rather than six constraints and includes a new ISR treatment.  

\subsubsection{Event Selection}
\label{section:qqqqselect}

A sample of hadronic events was selected by requiring more than 13 charged particles and a total visible energy exceeding $1.15~\Ebeam$.

$\qqgam$ events were suppressed by demanding an effective centre-of-mass
 energy~\cite{SPRIM}, after ISR emission, of greater than $161~\GeV$. The algorithm for assessing the $\ee$ collision energy considers both the emission of unobserved ISR photons in the beam-pipe and photon candidates detected in the electromagnetic calorimeters.

The \DURHAM\ jet clustering algorithm  ~\cite{DURHAM} with $y_{cut}$ of $0.002$ was applied to the event. If any of the resulting jets contained less than three particles or had an invariant mass smaller than $1~\GeVm$, clustering was continued to a higher value of $y_{cut}$. Events with less than four jets were then rejected, while events containing six or more jets were re-clustered into five objects representing four quarks plus one hard gluon jet.

Events containing b-quarks were rejected using the \DELPHI\ b-tag package \cite{aabtag}, this removes 17$\%$ of $\ZZ$ events and 6$\%$ of $\qqgam$ background while reducing the signal efficiency by only 0.2$\%$ . A four constrained kinematic fit \cite{delpaper183} was applied to the remaining events, enforcing conservation of energy and momentum. 

A variable to discriminate between $\qq$ events with hard gluon radiation and signal events was constructed. This compound variable relied upon the fitted jet energies and the inter-jet angles. The expected fraction of $\qqqq$ events in the selected sample, the event purity, was parametrised as a function of this variable. Events with an estimated purity below 25$\%$ were rejected.

The performance of the event selection is shown in Tab. \ref{tab:evtsel}.  Further information on a selection of $\WW$ events in \DELPHI\ with a similar performance is available in \cite{189xsec}.

\subsubsection{Likelihood Function}

For each of the selected events an event likelihood was constructed :
\begin{eqnarray}
 {\cal L}_e (\mw,\Gamma_{\rm W}) = \int \!\! \int \sum_i w_{i,e} \cdot
p_{i,e} (m_x,m_y)  \cdot  \\ \left[ P_e^{eff}\cdot \right. \nonumber  
\left. S(m_x,m_y,\mw,\Gamma_{\rm W}) + (1-P_e^{eff})\cdot B \right] dm_x dm_y.
\end{eqnarray}

\noindent As in \cite{delpaper183} the signal function  $S(m_x,m_y,\mw,\Gamma_{\rm W})$ consists of Breit-Wigner terms for the WW and the $\ZZ$ contribution and a phase space correction factor.  A flat distribution $B$ accounts for background processes and wrong jet pairings in the signal events. Both $S$ and $B$ were normalised to 1 over the integration area. The fraction of the signal and background likelihoods used for each event depend upon the event purity $P_e^{eff}$. This purity was parametrised as a function of a discriminating variable as described above.  

The sum $\sum_i w_{i,e} \cdot p_{i,e} (m_x,m_y)$ is a weighted sum of the probability densities $p_{i,e}$  that the event $e$ corresponds to two heavy objects
with mass $m_x$ and $m_y$. 

The probability density $p_{i,e}(m_x,m_y) \propto \exp\left[-\frac{1}{2}\chi^2_{i,e}(m_x,m_y)\right]$  was determined for all jet pairings (three possibilities for a four-jet event and ten for a five-jet event) and with three different clustering  algorithms (\DURHAM~\cite{DURHAM}, \CAMJET~\cite{CAMJET} and 
\DICLUS~\cite{DICLUS}). The relative probabilities $w_{i,e}$ that the 
corresponding jet pairing was the correct one were estimated using jet
charge information and the transverse momentum of the gluon candidate 
(see~\cite{delpaper183}). The three jet clustering algorithms were given the
 same weight.

The probability was calculated using a Gaussian approximation for the
$\chi^2$: 
\begin{equation} \chi^2_{i,e}(m_x,m_y) \approx
  \chi^2_{\rm 4C} + ({\bf m-m^{\rm fit}})^T {\bf V}^{-1}({\bf m-m^{\rm
  fit}}) \end{equation}
with 
$$ {\bf V} = \left( \begin{array}{cc}
                     \sigma^2_{m_x} & \sigma_{m_x}\sigma_{m_y}\rho_{xy} \\
                     \sigma_{m_x}\sigma_{m_y}\rho_{xy} & \sigma^2_{m_y}
                     \end{array}\right) \mbox{ , }
   {\bf m} = \left( \begin{array}{c}
                     m_x\\
                     m_y
                     \end{array}\right) \mbox{ and }
   {\bf m^{\rm fit}} = \left( \begin{array}{c}
                     m_x^{\rm fit}\\
                     m_y^{\rm fit}
                     \end{array}\right). \nonumber $$

\noindent The masses $m_x^{\rm fit}$, $m_y^{\rm fit}$, their errors $\sigma_{m_x}$ and
$\sigma_{m_y}$ and the correlation between them, $\rho_{xy}$, are taken
from a four constrained kinematic fit. When the  $\chi^2_{\rm 4C}$
 is larger than the number of degrees of freedom ($NDF=4$), the
$\chi^2_{i,e}(m_x,m_y)$ is rescaled with a factor $NDF
/ \chi^2_{\rm 4C}$ in order to compensate for non-Gaussian
resolution effects. This procedure decreases the computing time taken by an order of magnitude compared with the full six constrained fit~\cite{delpaper183}, while resulting in only a minimal reduction in the W mass precision obtained ($2 \pm 1 \%$).
 
A new feature of this analysis is a treatment of events under the collinear ISR hypothesis. A kinematic fit was performed with modified constraints to simulate the emission of an ISR photon of momentum $p_z$ inside the beam pipe:

\begin{equation} \sum_{i=1}^{n_{\rm objects}} (E, p_x, p_y, p_z)_i =
(\sqrt{s}-|p^{\rm fit}_z|,0,0,p^{\rm fit}_z). \end{equation}

The probability that the missing momentum in the $z$ direction was indeed due to an unseen ISR photon was extracted from the simulation as a function of $ |p^{\rm fit}_z| / \sigma_{p_z}$, where $\sigma_{p_z}$  is the estimated error on the fitted $z$ momentum component; only events with this ratio greater than 1.5 were treated with the mechanism described below.

\begin{figure}[ht]
 \epsfig{file=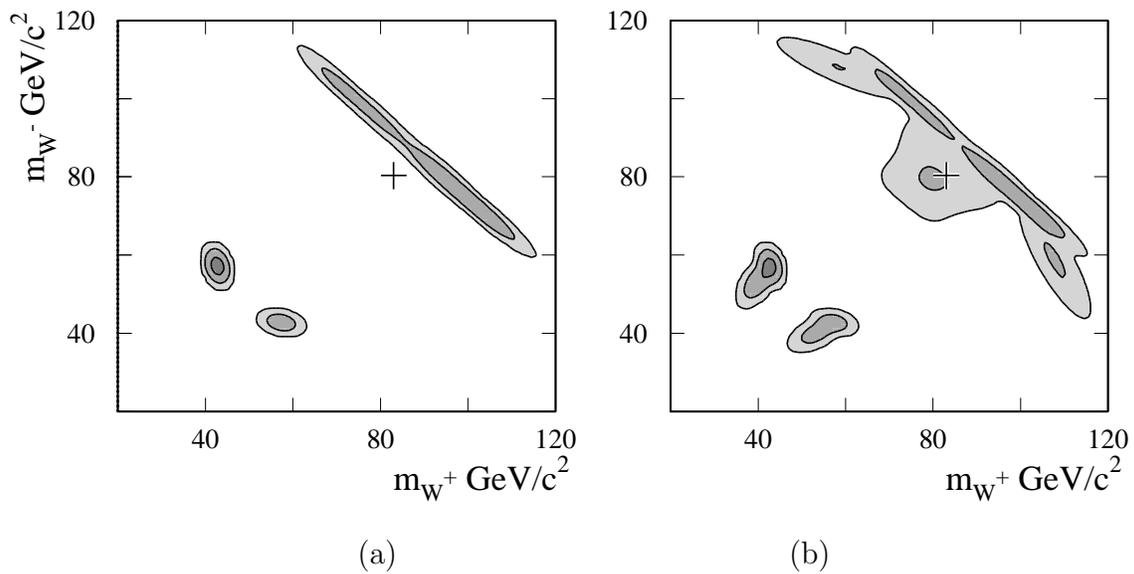,width=\textwidth}
 \begin{tabular}{cc}
 \hspace{5cm} (a) & \hspace{5cm}(b) \\
 \end{tabular}
  \caption{An example of the reconstructed probability density function $\sum_i w_{i,e} \cdot p_{i,e} (m_+,m_-)$  
(see text) for the invariant masses in a simulated 4-jet $\qqqq$ event without (a) and with (b) the hypothesis 
of collinear ISR. The first 3 sigma contours are shown. 
The normalization of the different solutions prevents the high mass
contours from reaching the 1 sigma probability level, while the small
difference in the low mass solutions originates from the jet charge information.
The generated masses of the two W bosons in the event are marked with a cross.}
  \label{fig:ISR}
 \end{figure}

Additional probability density $p_{i,e}$ terms were then included in the likelihood sum for these events, with a relative weight factor derived from the ISR hypothesis probability. An example of the effect of including the ISR hypothesis is shown in Fig. \ref{fig:ISR}. This treatment was applied to $16\%$ of the events and resulted in an improvement of the expected W mass error for these events of $15\%$.  

\subsection{Mass and Width Extraction}
\label{sec:massextrac}

\begin{figure}[ht]
 \begin{tabular}{cc}
 \epsfig{file=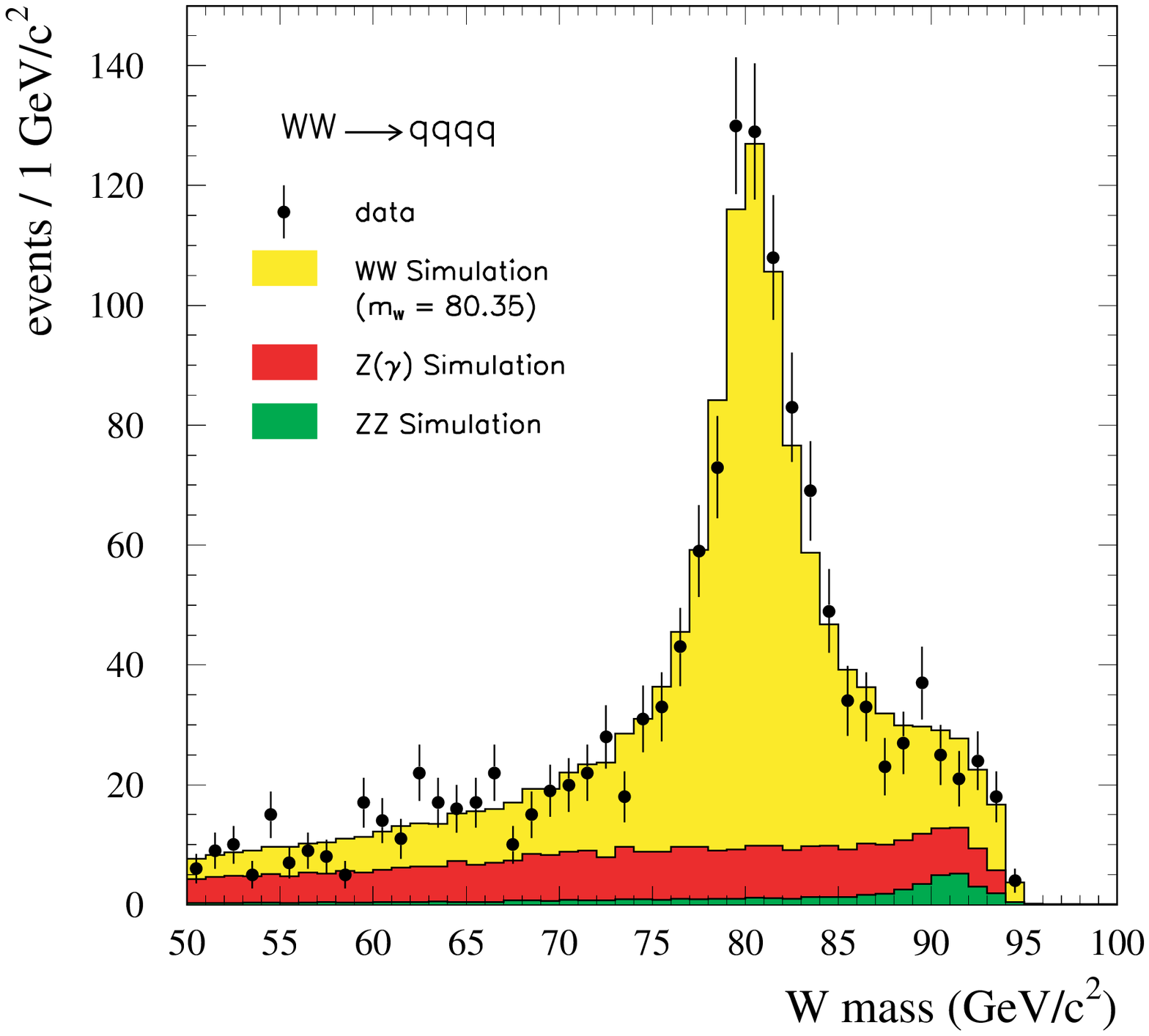,width=0.5\textwidth} & 
  \epsfig{file=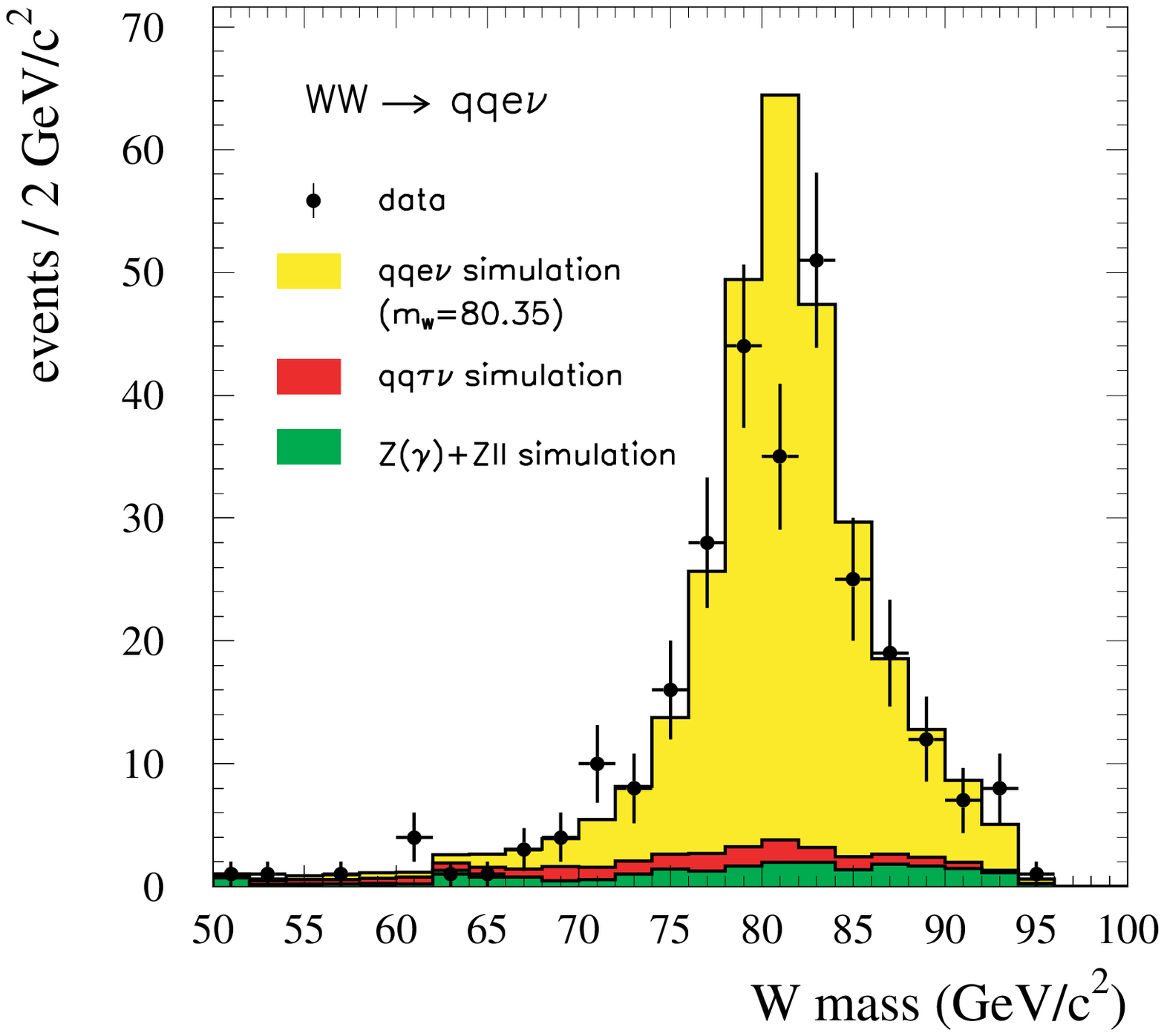,width=0.5\textwidth} 
\\
(a) & (b) \\
  \epsfig{file=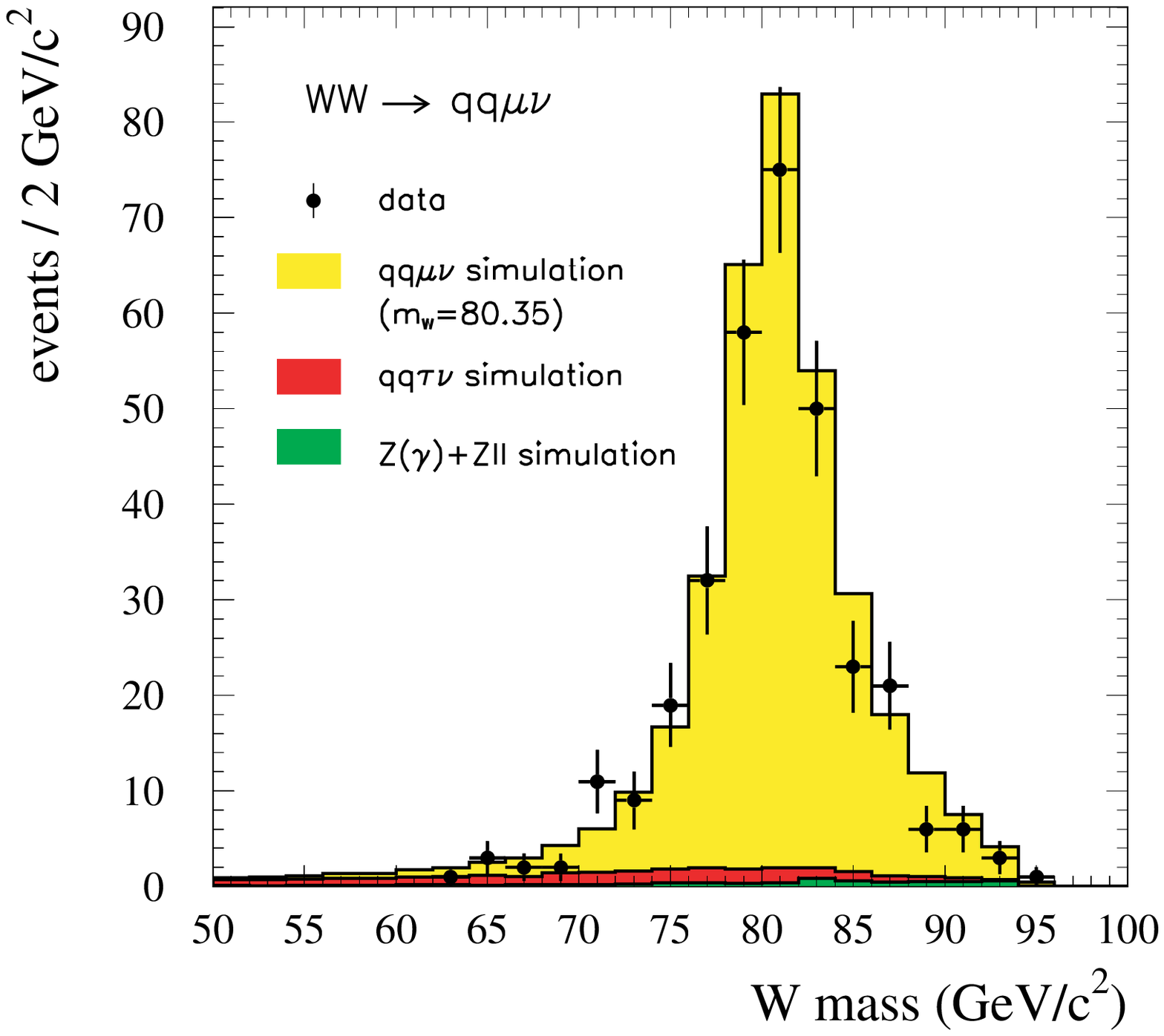,width=0.5\textwidth} &  \epsfig{file=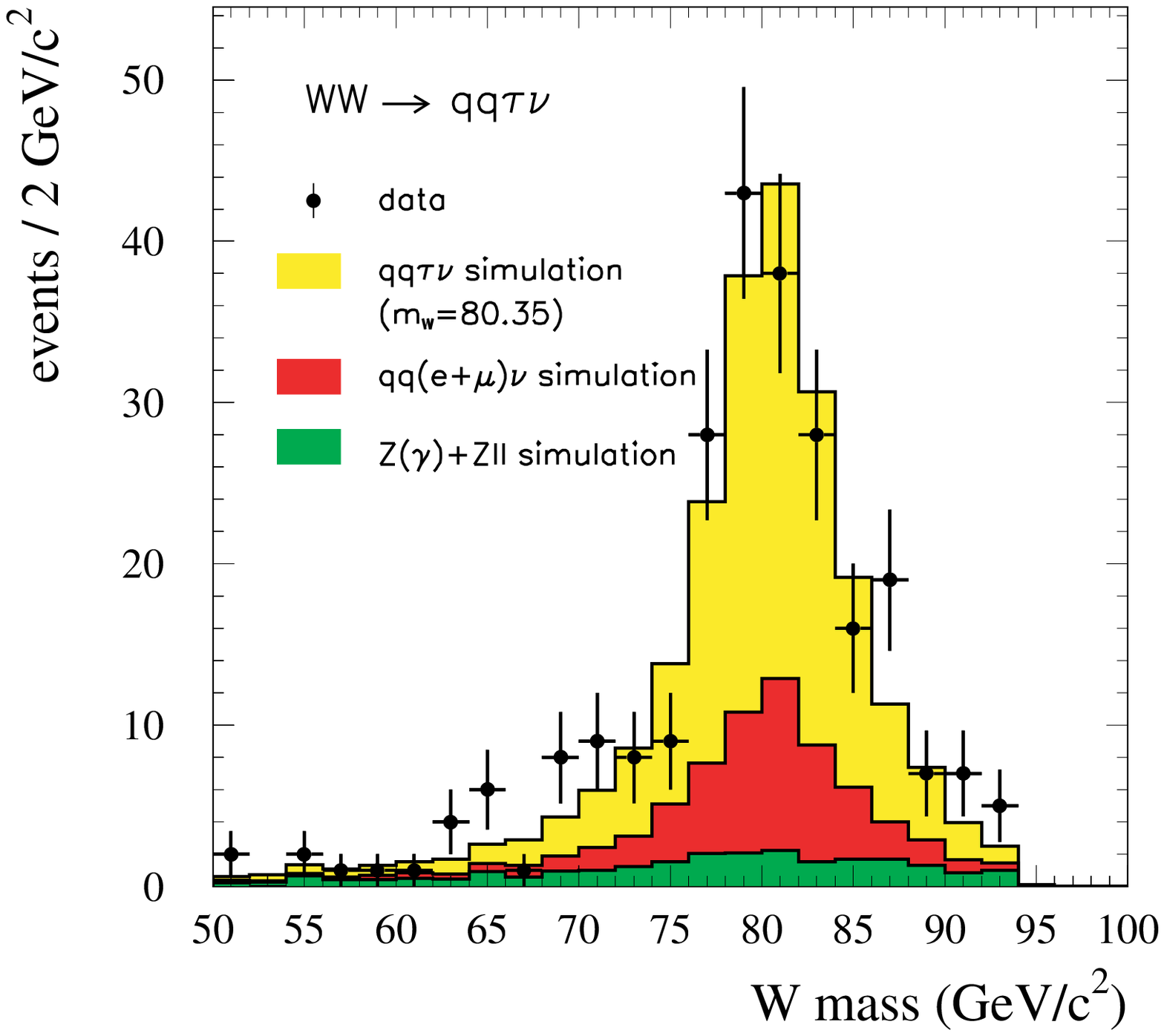,width=0.5\textwidth}
\\
(c) & (d) \\
 \end{tabular}
  \caption{The distribution of the reconstructed W masses from a kinematic fit with five constraints imposed in the (a) $\qqqq$ , (b) $\enqq$, (c) $\mnqq$ and (d) $\tnqq$ analysis channels. In the $\qqqq$ channel, only the jet pairing with the highest probability is included in this figure.}
  \label{fig:mass}
 \end{figure}

The distribution of the reconstructed invariant masses of the selected events after applying a kinematic fit, imposing four-momentum conservation and the equality of the two di-jet masses, are shown in Fig. \ref{fig:mass}. This plot is provided for illustrative purposes only, the mass and width fitting procedure is described below.

The combined likelihood of the data can be obtained from the product of the event likelihoods described above. The W mass and width were extracted from maximum likelihood fits. The W mass fit is performed assuming the standard model value for the W width.  The W width was obtained assuming a mass of $80.35~\GeVm$. The correlation between $\mw$ and $\gw$ was found to have a negligible impact on the extracted width value. 
 
The mass and width analyses have been calibrated separately in each of the decay channels ($\qqqq$,$\enqq$,$\mnqq$,$\tnqq$). The biases of the analyses were estimated by re-weighting generated simulation samples to obtain the fitted mass and width values. The re-weighting was performed using the extracted matrix element of the  \EXCALIBUR\ generator. The linearity of the mass analysis was estimated using independent simulation samples generated at three W mass values, while the re-weighting procedure was used for the width analysis.  The analyses were corrected with the calibration results, and the statistical error on the bias is included in the systematic error. 

The analyses were checked by performing fits to a large number of samples of simulation events. Each sample was comprised of a mixture of signal and background simulation events to represent the expected distribution in the data. The pull distribution $\frac{m_{fit}-m_W}{\sigma_{fit}}$ was demonstrated to be compatible with a Gaussian of width one to an accuracy of better than $1\%$. The mean expected statistical error in the W mass was $262~\MeVm$ for $\enqq$, $203~ \MeVm$ for $\mnqq$, $311~ \MeVm$ for $\tnqq$ and $104~ \MeVm$ for the $\qqqq$ channel. 

\section{Systematic Uncertainties}
\label{sec:syst}

The sources of systematic error that have been considered for the W mass and width determinations are described in the subsections below. The results of these studies are summarised in Tabs. \ref{tab:systmw} and \ref{tab:systgw}.

\subsection{Calibration}

The accuracy with which the bias of the analysis can be determined is limited by the size of the generated simulation samples. Sufficient events were generated to limit this error to less than $10\%$ of the statistical error in any given channel. The calibration procedure is described in Sec. \ref{sec:massextrac}.

\subsection{Detector Effects}

The data taken at the $\Z$ peak were used to study, and limit, possible errors in the detector simulation model.

Muon studies were performed on a selected sample of $\Z \rightarrow \mumu$ events. From the di-muon sample corrections to the inverse momentum scale, $1/p$, were calculated separately for positive and negative muons as a function of the lepton polar angle. The systematic error on this correction was estimated by varying it by half of its value. The momentum resolution (typically 0.001 in $1/p$) was found to be slightly better in simulation than in the data (a maximum difference of $10\%$). This was corrected by smearing the simulation with a Gaussian. An extra smearing of $0.0005$ in $1/p$ was used to estimate the systematic error coming from this correction. The combined systematic error from these corrections is quoted for the $\mnqq$ channel as the lepton correction systematic error in Tabs. \ref{tab:systmw} and \ref{tab:systgw}.
 
The correction of the energy scale of electrons was determined from Bhabha events at the Z peak in different polar angle regions. The residual systematic error on this absolute energy scale was estimated to be $0.5\%$. In each of these polar angle regions, the energy resolution of simulation events was degraded by applying a Gaussian smearing, and the residual error on this smearing was estimated to be $1\%$. The dependence of the energy calibration as a function of the electron energy was checked using low energy electrons from Compton events at the Z peak, and high energy electrons from radiative Bhabha scattering at high centre-of-mass energy. In these cases the true energy of the lepton was deduced from 3-body kinematics using only the angular information and assuming that the unseen particle was along the beam axis. The absolute energy calibration was found to be compatible with requiring no additional corrections in all energy ranges. The systematic error coming from this source was estimated assuming a $1\%$ change of slope in the energy calibration in the range of interest (between 25 and 70 $\GeV$). The lepton correction uncertainty for the $\enqq$ channel, quoted in Tabs. \ref{tab:systmw} and \ref{tab:systgw}, is the quadratic sum of the errors from these three error sources.

The lepton correction error is not quoted for the $\tnqq$ channel as the reconstructed tau lepton carries no information for the reconstructed W mass.

Jet energies were studied in $\Z \rightarrow \qq$ events as a function of the polar angle and reconstructed energy of the jet. The comparison of data and simulation showed agreement within a band of $\pm 2\%$ over most of the \DELPHI\ detector's angular coverage and an overall uncertainty of $1\%$ was estimated. The simulation was smeared by a Gaussian function to improve the description of the observed energy spread in the data and the residual error on this smearing was estimated to be $\pm 4\%$. 
The dependence of the energy calibration as a function of the jet energy was checked using low energy jets from $\qq + gluon$ events at the Z peak and high energy jets from radiative Z's at high centre-of-mass energy. The true jet energy
 was estimated from 3-body kinematics; in the radiative events the unseen photon was assumed to be along the beam axis. No additional energy calibration slope was necessary over the relevant energy range (25 to 75 $\GeV$) and a $1\%$ change in slope was used to calculate the systematic error.
A study of the acollinearity of jets in $Z \rightarrow \qq$ events was performed and an appropriate smearing to the simulation of the jet angular direction was estimated. A systematic error was estimated by applying an extra $5 ~\mrad$ angular smearing. The jet correction uncertainty, quoted in Tabs. \ref{tab:systmw} and \ref{tab:systgw}, is the quadratic sum of these four errors.
 
A possible source of angular distortion in \DELPHI\ is the uncertainty on the length to width ratio of the detector. The detector is aligned relative to the vertex detector, the largest uncertainty being the radius of this detector which is known to a precision of $\pm 0.1\%$. This error is listed in the $\mw$ systematics table under aspect ratio. The corresponding uncertainty for the width measurement is less than 10 $\MeV$.

\subsection{Background Description}

The background level was changed by $\pm 10\%$ in the simulation; this easily covers the expected uncertainty in the accepted cross-section, as discussed in \cite{189xsec}. The dominant background source, $\Z \rightarrow \qq (\gamma)$, was generated using both \JETSET\ and \HERWIG\ fragmentation models and mass fits performed using both these samples. The 4-jet rate in data and simulation was also studied using events collected at the $\Z$ peak.
It is concluded that the background description is a relatively small component of the systematic error for the mass measurement, as shown in Tab. \ref{tab:systmw} and somewhat more important for the width (see Tab. \ref{tab:systgw}).

\subsection{Fragmentation}

A study of the possible effects on $\mw$ and $\gw$ of the simulation of the event fragmentation was performed, the results are provided in Tabs. \ref{tab:fragmw} and \ref{tab:fraggw} .

Events were produced using a fast simulation package where the value of $\Lambda_{QCD}$ and $\sigma_{q}$ were changed with respect to the standard \DELPHI\ \JETSET\ tuned values \cite{deltune} by twice their estimated errors \footnote{The dominant systematic error components of the tuning uncertainties were estimated from a comparison of fits with a range of input data distributions}. The estimated errors are $\pm 0.018~\GeV$ for $\Lambda_{QCD}$ and $\pm 0.007~\GeV$ for $\sigma_{q}$.

The event samples generated by \ALEPH\ in the context of the LEP WW workshop are used to compare results from the \HERWIG\ and \JETSET\ fragmentation models. The \HERWIG\ events were produced with a recent tuning \cite{rudolph} which provides a better description of the data than previous \HERWIG\ versions.

The results on the \DELPHI\ tuning of \JETSET\ are also compared with those of the \ALEPH\ collaboration. This comparison cross-checks several effects as the \ALEPH\ events were produced with a different generator (\KORALW) from the \DELPHI\ events which includes a different ISR and final-state radiation (FSR) treatment.

As the results in Tabs. \ref{tab:fragmw} and \ref{tab:fraggw} are all compatible with zero, we quote a systematic error from fragmentation reflecting twice the statistical precision of the \JETSET\ tuning parameter studies in the combined semi-leptonic and fully-hadronic channels.

\subsection{Mixed Lorentz Boosted Z's}

The agreement between data and simulation can be cross-checked using the method of Mixed Lorentz Boosted Z's. Z events were selected from the Z peak data sample collected during 1998 and the corresponding simulation sample. Through a suitable choice of Lorentz boost and superimposing two Z events a WW event may be emulated. The angular distribution of the Z events used was chosen to match that expected in WW events. 

This technique has been applied to the $\qqqq$ mass measurement
and the hadronic system in semi-leptonic events. The differences between data and simulation obtained are $2 \pm 2 ~ \MeVm$ and $3 \pm 10 ~ \MeVm$ in the two topologies respectively, where the errors are statistical. These results reinforce the view that the quoted systematic errors for detector and fragmentation effects are  conservative. A study of possible intrinsic uncertainties of the MLBZ method \cite{mlbz} in the $\qqqq$ channel estimates the accuracy of this technique to be  $5 ~ \MeVm$, and demonstrates excellent agreement between data and simulation as a function of relevant event variables. 

\subsection{ISR}

A comparison of two independent models of ISR was performed. WW events produced with the \EXCALIBUR\ generator were re-weighted as a function of the total ISR energy in the event. The weights were obtained for each WW decay channel from a generator level study of the \KORALW\ \cite{koralw} ISR treatment (based on the YFS exponentiation approach) and that of the standard algorithm used in \DELPHI\ \QEDPS\ \cite{qedps} (based on a parton shower approach). The results of the W mass and width fits are shown in Tabs. \ref{tab:isrmw} and \ref{tab:isrgw}. We conservatively choose to quote the largest deviation observed in any of the channels. 

\subsection{LEP Beam Energy}

The average \LEP\ beam energy at \DELPHI\ is evaluated by the energy working group \cite{energy} at 15 minute intervals of running or after significant changes in the beam energy. The measured centre-of-mass energy is imposed as a constraint in the kinematic fit, and hence the relative error on the beam energy translates to the same fractional error on the W mass determination. The spread of  energies of the electrons and positrons in the LEP beams was found to have a negligible impact on the mass and width measurements.

\subsection{Bose-Einstein Correlations}

Bose-Einstein statistics dictate that the production amplitude for final state particles should be symmetrical under the exchange of identical bosons. The omission of these correlations between particles from different W bosons in our standard simulation could lead to a systematic error on the W mass and width measurement in the $\qqqq$ channel. 
A clear picture has yet to emerge from the experimental study of this phenomenon \cite{bose}.

To evaluate the possible size of the effect on the mass and width measurements
we have considered several phenomenological models; the results are given in Tab. \ref{tab:systbe}. The relevant value for the systematic uncertainty is the difference between the shifts obtained from Bose-Einstein correlations inside individual W's and that between W's. The models used are:
\begin{itemize}
 \item The \LUBOEI\ algorithm \cite{luboei} in \JETSET\ changes the momentum vectors  of identical final state bosons to model the two particle correlation 
   function, and then offers a range of options to restore energy and momentum 
   conservation. 
 \item Global re-weighting aims to reproduce the enhancement of
   identical bosons close in phase-space by giving weights to
   events. This procedure does ensure energy and momentum
   conservation, but may adversely affect other event distributions.
   \DELPHI\ results were obtained using the Kartvelishvili/Kvatadze/M{\o}ller 
   (KKM) re-weighting scheme \cite{kkm} and reported on in \cite{delpaper183}.
 \item In \cite{delpaper183} we also reported results for a study based on a 
   modification of the \JETSET\ fragmentation model introducing quantum 
   mechanical 2-particle and 3-particle interference (ST) for 
   identical bosons using a local re-weighting technique \cite{sharka}.
\end{itemize}

The version of \LUBOEI\ model studied here is $BE_{32}$, in which a local energy and momentum conservation procedure is applied \cite{luboei}.
Six sets of fully simulated Z events were generated at a range of values of correlation strength ($\lambda$) and radius ($r$), and the four momentum difference, $Q$, between all selected same charge particle pairs was calculated. An interpolation was performed on the basis of the $Q$ distribution to obtain $\lambda =1.35$ and $r=0.6~\ufm~(PARJ(93)=0.34 ~\GeV)$ which provide the optimal description of the \DELPHI\ Z peak data. The simulated event samples were produced with the standard \DELPHI\ \JETSET\ tuning. The W mass and width shifts were evaluated using samples produced with fast simulation.

\subsection{Colour Reconnection}

The hadronization of two W bosons in the $\qqqq$ channel may not occur 
independently. The colour flow between the W's may lead to a shift in 
the measured W mass. Although experimental work is progressing \cite{L3Colour}, a suitable sensitivity has not yet been reached by the measurements to limit the effect on $\mw$ .

In the previous publication \cite{delpaper183} we reported results with the
\ARIADNE\ colour reconnection model, the results are repeated in Tab. \ref{tab:cr}. However, we note that the current version of the \ARIADNE\ 2 model is disfavoured by \LEP1\ data \cite{opalariadne}, and that we do not consider the \ARIADNE\ 3 model for the systematic error assessment as it allows perturbative phase reconnection where calculations have shown the effect to be small \cite{perturb}.

The Sj\"ostrand/Khoze models of colour reconnection are available in the \JETSET\ frame-work. We have used the SK1 and SK2 models. Results for the SK1 model are quoted for 30$\%$ of reconnected events, this is the same reconnected fraction as in the SK2 model. Events were produced using \DELSIM\ and a fast detector simulation. By processing the same event sample through the full detector simulation and the fast simulation, the reliability of the fast detector simulation for this study was clearly demonstrated. The simulated event samples were produced with the standard \DELPHI\ \JETSET\ tuning.

However, we report that the standard implementation of SK1 shows 
numerical instabilities that reduce the reliability of the
model used. In this model the reconnection probability is a function of the 
string overlap. This overlap is calculated by numerical integration through
sampling. The accuracy of this calculation is improved by using 1000 sampling points rather than the default value of 100. In addition, we also report results from a more efficient sampling of the string overlap, in which the sampling is performed along strings taking into account their life-time and the total overlap is calculated as the sum of the overlap of pairs of string pieces.

\begin{figure}[htbp]
  \includegraphics[height=8cm]{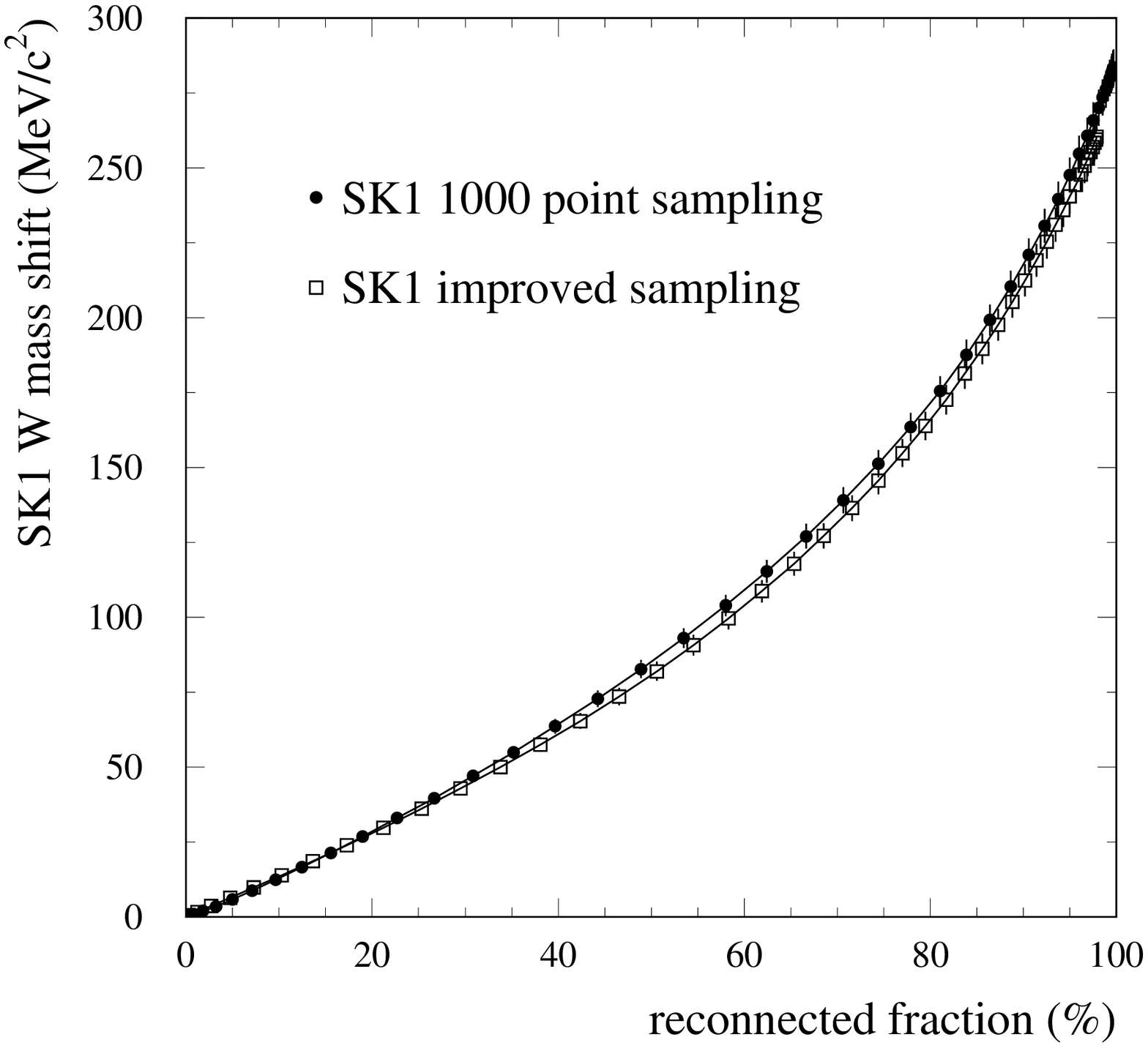} 
  \includegraphics[height=8cm]{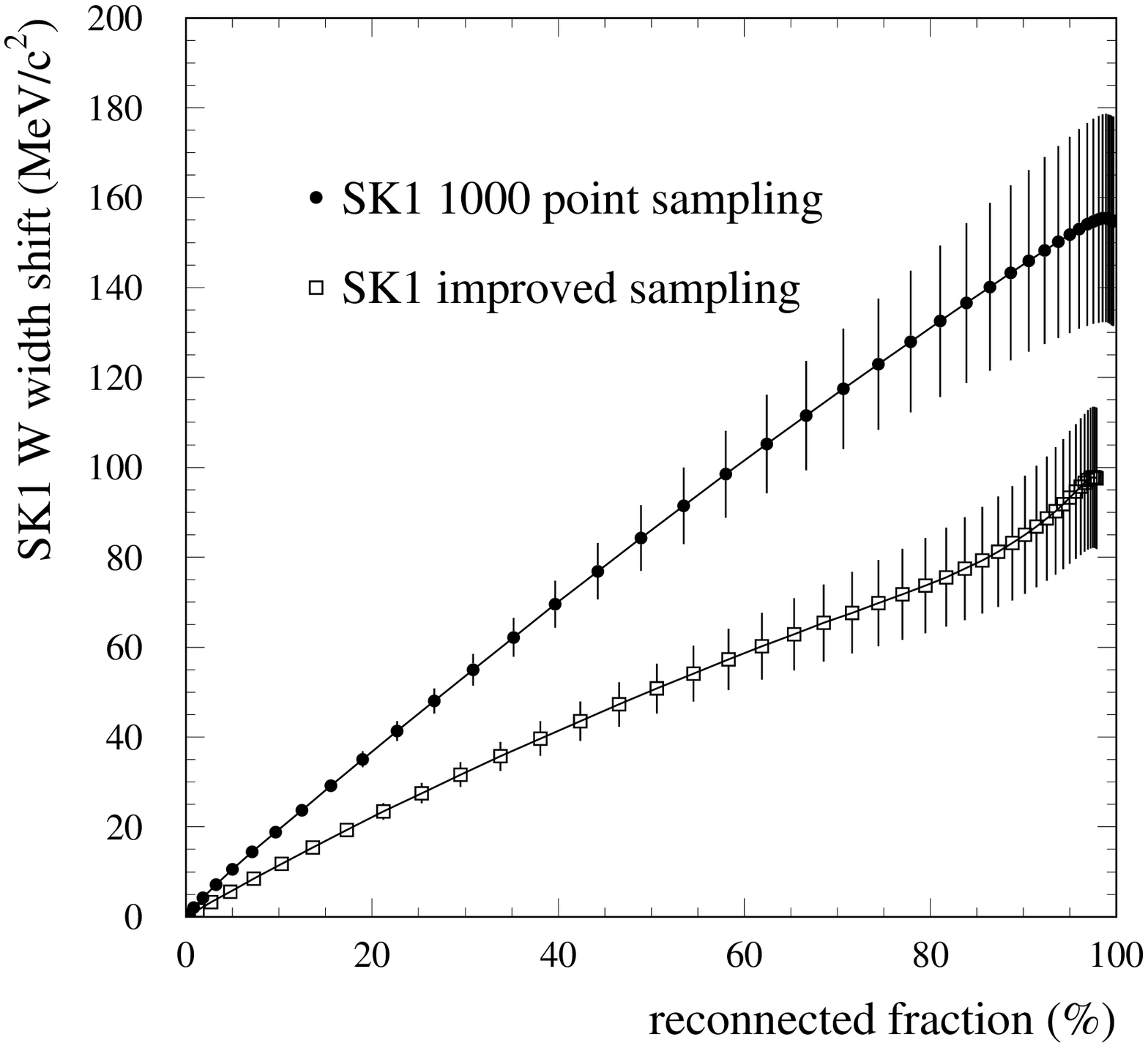} 
  \caption{Observed shift in the fitted mass (left) and width (right)
  as a function of the fraction of reconnected events, using the
  SK1 implementations as described in the text.}
  \label{fig:reco}
 \end{figure}

The observed W mass and width shifts are given in Tab. \ref{tab:cr} and in Fig. \ref{fig:reco} the observed W mass and width shifts are shown as functions of the reconnected fraction of events in the SK1 model.

We conservatively choose to quote as a systematic error the largest effect observed in our studies; this is approximately $50~\MeV$ for both the mass and width analyses. 

\subsection{Correlations}

The components of the systematic error arising from the jet energy scale corrections, aspect ratio, ISR, fragmentation and LEP beam energy are taken as correlated between the analyses in the different decay channels for the 189 GeV data. The background description and lepton modelling uncertainties are treated as uncorrelated between WW decay channels.

In Sec. \ref{sec:comb2} a combination with the previously published \DELPHI\ results is performed. The LEP energy correlation matrix is used \cite{energy} in this process. The calibration statistics error is uncorrelated between years, while all other systematic errors are conservatively assumed to be fully correlated between years.

\section{Results}
\label{sec:res}
\subsection{189  {\boldmath $\GeV$ } Results}

The W mass and width results of the analyses described in this paper are presented in Tabs. \ref{tab:res189mw} and \ref{tab:res189gw}. The error is divided into its statistical component, indicated ($stat$), the main systematic component ($syst$) and the systematic from the beam energy uncertainty ($LEP$). In the $\qqqq$ channel an error from final state interaction effects ($FSI$) is also included.

The semi-leptonic and fully-hadronic W mass results are combined and the following result obtained:
 \begin{eqnarray*}
   \mw & = & 80.387 \pm 0.087 (stat) \pm 0.034 (syst) \pm 0.017 (LEP) \pm 0.035 (FSI) ~ \GeVm .
 \end{eqnarray*}

\noindent The combined result for the W width is :
 \begin{eqnarray*}
   \gw & = & 2.205 \pm 0.195 (stat) \pm 0.059 (syst) \pm 0.047 (FSI) ~ \GeVm .
 \end{eqnarray*}

\subsection{Combined {DELPHI} Results}
\label{sec:comb2}
These results are combined with the previously published \DELPHI\ results on the W mass determination by direct reconstruction \cite{delpaper172,delpaper183} and the result from the threshold cross-section measurement \cite{delpaper161} to obtain:
\begin{eqnarray*}
   \mw & = & 80.359 \pm 0.074 (stat) \pm 0.032 (syst) \pm 0.017 (LEP) \pm 0.033 (FSI) ~ \GeVm .
 \end{eqnarray*}

\noindent This combination has a $\chisq$ of 4.5 with 6 degrees of freedom.
The combined result on the W mass from the $\qqqq$ channel alone is :
\begin{eqnarray*}
   \mw & = &  80.369 \pm  0.091 (stat) \pm  0.029 (syst) \pm 0.017 (LEP) \pm 0.056 (FSI) ~ \GeVm
 \end{eqnarray*}

\noindent and for the $\lnqq$ channel alone is :
\begin{eqnarray*}
   \mw & = & 80.327 \pm 0.128 (stat) \pm 0.045 (syst) \pm 0.017 (LEP) ~ \GeVm .
 \end{eqnarray*}

The mass difference between the measurements for the fully-hadronic and semi-leptonic channels, $\Delta\mw(\qqqq-\lnqq)$, has been determined:
\begin{eqnarray*}
 \Delta\mw(\qqqq-\lnqq) =  39 \pm 159~\MeVm.    
\end{eqnarray*}

A significant non-zero value for $\Delta\mw$ could indicate that 
FSI effects are biasing the value of $\mw$ determined from the $\qqqq$
 events.
As $\Delta\mw$ is primarily of interest as a check of the possible 
effects of final state interactions, the errors from FSI effects are set to 
zero in this determination: all other errors and correlations were as described above.

The combination for the W width with the 1997 data \cite{delpaper183} yields :
\begin{eqnarray*}
   \gw & = & 2.266 \pm 0.176 (stat) \pm 0.056 (syst) \pm 0.052 (FSI) ~ \GeVm .
 \end{eqnarray*}


\subsection*{Acknowledgements}
\vskip 3 mm
 We are greatly indebted to our technical 
collaborators, to the members of the CERN-SL Division for the excellent 
performance of the LEP collider, the LEP energy working group for the beam energy estimation and to the funding agencies for their
support in building and operating the DELPHI detector.\\
We acknowledge in particular the support of \\
Austrian Federal Ministry of Science and Traffics, GZ 616.364/2-III/2a/98, \\
FNRS--FWO, Flanders Institute to encourage scientific and technological 
research in the industry (IWT), Belgium,  \\
FINEP, CNPq, CAPES, FUJB and FAPERJ, Brazil, \\
Czech Ministry of Industry and Trade, GA CR 202/96/0450 and GA AVCR A1010521,\\
Danish Natural Research Council, \\
Commission of the European Communities (DG XII), \\
Direction des Sciences de la Mati$\grave{\mbox{\rm e}}$re, CEA, France, \\
Bundesministerium f$\ddot{\mbox{\rm u}}$r Bildung, Wissenschaft, Forschung 
und Technologie, Germany,\\
General Secretariat for Research and Technology, Greece, \\
National Science Foundation (NWO) and Foundation for Research on Matter (FOM),
The Netherlands, \\
Norwegian Research Council,  \\
State Committee for Scientific Research, Poland, 2P03B06015, 2P03B11116 and
SPUB/P03/DZ3/99, \\
JNICT--Junta Nacional de Investiga\c{c}\~{a}o Cient\'{\i}fica 
e Tecnol$\acute{\mbox{\rm o}}$gica, Portugal, \\
Vedecka grantova agentura MS SR, Slovakia, Nr. 95/5195/134, \\
Ministry of Science and Technology of the Republic of Slovenia, \\
CICYT, Spain, AEN96--1661 and AEN96-1681,  \\
The Swedish Natural Science Research Council,      \\
Particle Physics and Astronomy Research Council, UK, \\
Department of Energy, USA, DE--FG02--94ER40817.

\pagebreak

\pagebreak

\begin{table}
 \begin{center}
  \begin{tabular}{|l|c|c|c|c|c|}
\hline
                    & \multicolumn{5}{|c|}{Event Selection} \\
Event Type          &  $\enqq$     & $\mnqq$   & $\tnqq$   & $\lnqq$       & $\qqqq$ \\
\hline \hline
$\enqq$             &    $259.5 $  & $0.3 $    & $43.6 $   & $303.4 $   & $4.2 $ \\
$\mnqq$             &    $0.5 $    & $319.6 $  & $10.8 $   & $330.9 $   & $2.5 $ \\
$\tnqq$             &    $12.1 $   & $13.4 $   & $155.4 $  & $180.9 $   & $8.0 $ \\
$\qqqq$             &    $0.2 $    & $0.2 $    & $1.7 $    & $2.1 $   & $1112.7 $ \\
Other 4f            &    $3.4 $    &  $3.0 $   & $4.4 $    &  $10.8 $   & $0.0 $\\
$\qqgam$ and other 2f & $8.0 $     &  $0.7 $   & $15.8 $   &  $24.5 $   & $346.9 $ \\
Total               & $283.7$      & $337.2 $  & $231.7 $  & $852.6$    & $1474.2 $ \\
\hline                                          
Data                &   $244$      &  $307$    &  $236$       &  $787$        &  $1481$ \\ \hline
  \end{tabular}
\caption{ Number of selected events from the 189 $\GeV$ data sample, and the corresponding number of expected events from the simulation. Column four is the sum of the three previous columns.}
  \label{tab:evtsel} 
 \end{center}
\end{table}


\begin{table}
\begin{center}
\begin{tabular}{|l|c|c|c|c|}
\hline
 \multicolumn{5}{|c|}{$\mw$ Fragmentation study ($\MeVm$)} \\
\hline
Event Type                & $\enqq$ & $\mnqq$ & $\tnqq$ & $\qqqq$ \\ 
\hline \hline
$\Lambda_{QCD} -2 \sigma$ & $0 \pm 14$  & $+9 \pm 16$  & $+30 \pm 30$  & $+3 \pm 6$ \\
$\Lambda_{QCD} +2 \sigma$ & $-4 \pm 14$ & $+10 \pm 16$ & $-10 \pm 30$ & $-7 \pm 6$  \\
$\sigma_{q} -2 \sigma$    & $0 \pm 14$  & $+30 \pm 16$ & $-3 \pm 30$  & $-9 \pm 6$ \\
$\sigma_{q} +2 \sigma$    & $0 \pm 14$  & $+30 \pm 16$ & $+27 \pm 30$  & $-3 \pm 6$  \\
\hline
$\ALEPH\ \HERWIG$ $-$ $\ALEPH\ \JETSET$ & $75 \pm 52$ &$8 \pm 42$ & $106 \pm 66$&  $-6 \pm 18$ \\
$\ALEPH\ \JETSET$ & $15 \pm 51$ &  $-17 \pm 42$ & $-81 \pm 65$ & $+3 \pm 18$\\
\hline
\end{tabular}
\caption{Results of a study of fragmentation effects on the $\mw$ measurement (see text). All results are given for the fitted mass analyses on the samples studied with respect to the standard \DELPHI\ \JETSET\ sample, unless otherwise stated. The statistical error on the observed difference is also given.}
\label{tab:fragmw}
\end{center}
\end{table}

\begin{table}
 \begin{center}
\begin{tabular}{|l|c|c|c|c|}
\hline
\multicolumn{5}{|c|}{$\gw$ Fragmentation study ($\MeVm$)} \\
\hline 
Event Type                & $\enqq$ & $\mnqq$ & $\tnqq$  & $\qqqq$ \\ 
\hline \hline
$\Lambda_{QCD} -2 \sigma$ & $-5 \pm 30$  & $-46 \pm 35$ & $-8 \pm 59$   & $-4 \pm 12$\\
$\Lambda_{QCD} +2 \sigma$ & $+11 \pm 30$ & $-47 \pm 35$ & $+57 \pm 59$  & $+15 \pm 12$ \\
$\sigma_{q} -2 \sigma$    & $+16 \pm 30$  & $-76 \pm 35$ & $-11 \pm 59$ & $-11 \pm 12$\\
$\sigma_{q} +2 \sigma$    & $+28 \pm 30$  & $-78 \pm 35$ & $-30 \pm 59$ & $-2 \pm 12$ \\
\hline
$\ALEPH\ \HERWIG$ - $\ALEPH \JETSET$ & $ -148 \pm 113$  & $-47 \pm 95$ & $ -155 \pm 158 $& $-68 \pm 38 $  \\
$\ALEPH\ \JETSET$  &  $+264 \pm 108$ & $-30 \pm 95$ & $+176 \pm 157$  & $+4  \pm 43$ \\
\hline
\end{tabular}
\caption{Results of a study of fragmentation effects on the $\gw$ measurement (see text). All results are given for the fitted mass analyses on the samples studied with respect to the standard \DELPHI\ \JETSET\ sample, unless otherwise stated. The statistical error on the observed difference is also given.}
\label{tab:fraggw}
 \end{center}
\end{table}

\begin{table}
\begin{center}
\begin{tabular}{|l|c|c|c|c|}
\hline
\multicolumn{5}{|c|}{$\mw$ ISR study ($\MeVm$)} \\
\hline
Model                & $\enqq$ & $\mnqq$ & $\tnqq$  & $\qqqq$ \\ 
\hline \hline
\KORALW\ re-weighted - \QEDPS\ & $+2 \pm 7$ & $0 \pm 5$ & $-12 \pm 8$  & $-16 \pm 3$ \\
\hline
\end{tabular}
\caption{A comparison of two ISR treatments for the $\mw$ measurement, the \KORALW\ form and the \QEDPS\ treatment. The statistical error on the result is also provided.}
\label{tab:isrmw}
 \end{center}
\end{table}

\begin{table}
 \begin{center}
\begin{tabular}{|l|c|c|c|c|}
\hline
\multicolumn{5}{|c|}{$\gw$ ISR study ($\MeVm$)} \\
\hline
Model                & $\enqq$ & $\mnqq$ & $\tnqq$ &  $\qqqq$ \\ 
\hline \hline
\KORALW\ re-weighted - \QEDPS\ & $+13 \pm 13$ & $+14 \pm 10$ & $+4 \pm 16$ & $-16 \pm 5$ \\
\hline
\end{tabular}
\caption{A comparison of two ISR treatments for the $\gw$ measurement, the \KORALW\ form and the \QEDPS\ treatment. The statistical error on the result is also provided.}
\label{tab:isrgw}
\end{center}
\end{table}


  \begin{table}
 \begin{center}
    \begin{tabular}{|l|c|c|} \hline
\multicolumn{3}{|c|}{Bose-Einstein Correlations Study ($\MeVm$)} \\
\hline
      Model & $\mw$ shift & $\gw$ shift \\ \hline \hline
      KKM model ($183 \GeV$) &  -10 $\pm$ 10   & - \\
     \hline
      ST model $BE$ ($183 \GeV$) inside W's - none  & +0 $\pm$ 10 & - \\
      ST model $BE$ ($183 \GeV$) between W's - inside W's & +3 $\pm$ 11  & - \\
     \hline
      LUBOEI $BE_{32}$ ($189 \GeV$) inside W's - none &  +18 $\pm$ 5 &   +13 $\pm$  11 \\
      LUBOEI $BE_{32}$ ($189 \GeV$) between W's - inside W's &  -32 $\pm$ 4 & +26 $\pm$  8  \\
    \hline
    \end{tabular}
    \caption{Results of the \DELPHI\ studies on Bose-Einstein Correlations, see text for details. The $BE_{32}$ samples were produced with $\lambda=1.35, r=0.6~\ufm$}
    \label{tab:systbe}
    \end{center}
    \end{table}


  \begin{table}[ht]
  \begin{center}
    \begin{tabular}{|l|c|c|} \hline
\multicolumn{3}{|c|}{Colour Reconnection Study ($\MeVm$)} \\
     \hline
      Model & $\mw$ shift & $\gw$ shift  \\
     \hline \hline
      \ARIADNE\ CR 2 ($183 \GeV$) & +28 $\pm$ 6 & -\\
      \ARIADNE\ CR 3 ($183 \GeV$) & +55 $\pm$ 6 & -\\ 
      \hline
      \JETSET\ SK1 & 46 $\pm$ 2 & 54 $\pm$ 3 \\
      \JETSET\ SK1 improved sampling & 44 $\pm$ 2 & 32 $\pm$ 3 \\
      \JETSET\ SK2 & -2 $\pm$ 5 &  37 $\pm$ 10 \\
      \hline
    \end{tabular}
      \caption{Results of the \DELPHI\ studies on Colour reconnection effects, see text for details. For the SK1 model results are given for 30$\%$ of reconnected events.}
      \label{tab:cr}
    \end{center}
    \end{table}


 \begin{table}[ht]
 \begin{center}
  \begin{tabular}{|l|ccc|c|c|}     \hline
\multicolumn{6}{|c|}{$\mw$  Systematic Errors ($\MeVm$)} \\ \hline 
Sources of systematic error & $\enqq$ & $\mnqq$ & $\tnqq$ & $\lnqq$ & $\qqqq$ \\ \hline \hline
 Statistical error on calibration & 18  & 15 & 23 & 10 & 7  \\
 Lepton corrections               & 29  & 11 & -  & 10 & -  \\
 Jet corrections                  & 39  & 27 & 48 & 35 & 18 \\
 Aspect Ratio                     & 2   & 2  & 2  & 2  & 4 \\
 Background                       & 10  & 3  & 4  & 3  & 5   \\
 Fragmentation                    & 20  & 20 & 20 & 20 & 12 \\
 I.S.R.                           & 16  & 16 & 16 & 16 & 16 \\ 
\hline
 LEP energy                       & 17  & 17 & 17 & 17 & 17 \\
\hline
 Colour reconnection              & -   & -  & -   &  - & 46  \\
 Bose Einstein correlations       & -   & -  & -   &  - & 32  \\ \hline
  \end{tabular}
  \caption{Contributions to the systematic error on the W mass measurement.}
  \label{tab:systmw}
 \end{center}
 \end{table}

\begin{table}[ht]
 \begin{center}
  \begin{tabular}{|l|ccc|c|c|}     \hline
\multicolumn{6}{|c|}{$\gw$  Systematic Errors ($\MeVm$)} \\ \hline
Sources of systematic error & $\enqq$ & $\mnqq$ & $\tnqq$ & $\lnqq$ & $\qqqq$ \\ \hline \hline
 Statistical error on calibration & 45 & 37 & 56 & 26 & 17 \\
 Lepton corrections         &  41 & 46 & -   & 28 & - \\
 Jet corrections            &  82 & 43 & 102 & 63 & 26\\
 Background                 &  29 &  8 & 82  & 19 & 40 \\
 Fragmentation              &  42 & 42 & 42  & 42 & 24 \\
 I.S.R.                     &  16 & 16 & 16  & 16 & 16 \\
\hline
 Colour reconnection        & -         & -  & -   &  - &  54   \\
 Bose Einstein correlations & -         & -  & -   &  - &  26  \\ \hline
  \end{tabular}
  \caption{Contributions to the systematic error on the W width measurement.}
  \label{tab:systgw}
 \end{center}
 \end{table}

\begin{table}[ht]
 \begin{center}
  \begin{tabular}{|l|ccccc|}     \hline
\multicolumn{6}{|c|}{189 $\GeV$ $\mw$ results ($\GeVm$)} \\ \hline 
Channel & $\mw$ & $stat.$ & $syst.$   & $LEP$     & $FSI$ \\
\hline \hline
$\enqq$ &  $80.478$ & $\pm 0.291$ & $\pm 0.059$ & $\pm 0.017$ & - \\
$\mnqq$ &  $80.195$ & $\pm 0.213$ & $\pm 0.042$ & $\pm 0.017$ & - \\ 
$\tnqq$ &  $80.114$ & $\pm 0.319$ & $\pm 0.059$ & $\pm 0.017$ & - \\
\hline
$\lnqq$ &  $80.253$ & $\pm 0.151$ & $\pm 0.046$ & $\pm 0.017$ & - \\
\hline
$\qqqq$ &  $80.466$ & $\pm 0.106$ & $\pm 0.028$ & $\pm 0.017$ & $\pm 0.056$ \\
\hline
  \end{tabular}
  \caption{189 $\GeV$ $\mw$ results. The error is divided into its statistical component, indicated $stat$, the main systematic component $syst$ and the systematic from the beam energy uncertainty $LEP$. In the $\qqqq$ channel an error from final state interaction effects $FSI$ is also included. The $\lnqq$ results represents the combination of the results obtained in the three semi-leptonic channels.}
  \label{tab:res189mw}
 \end{center}
 \end{table}

\begin{table}[ht]
 \begin{center}
  \begin{tabular}{|l|cccc|}     \hline
\multicolumn{5}{|c|}{189 $\GeV$ $\gw$ results ($\GeVm$)} \\ \hline 
Channel & $\gw$ & $stat.$ & $syst.$  & $FSI$ \\
\hline \hline
$\enqq$ & $4.358$ & $\pm 0.956$ & $\pm 0.115$ & - \\
$\mnqq$ & $2.353$ & $\pm 0.552$ & $\pm 0.086$ & - \\ 
$\tnqq$ & $2.799$ & $\pm 0.927$ & $\pm 0.149$ & - \\
\hline
$\lnqq$ & $2.842$ & $\pm 0.425$ & $\pm 0.088$ & - \\
\hline
$\qqqq$ & $2.025$ & $\pm 0.220$ & $\pm 0.058$ & $\pm 0.060$ \\
\hline
  \end{tabular}
  \caption{189 $\GeV$ $\gw$ results. The error is divided into its statistical component, indicated $stat$, the main systematic component $syst$ and the systematic from the beam energy uncertainty $LEP$. In the $\qqqq$ channel an error from final state interaction effects $FSI$ is also included. The $\lnqq$ results represents the combination of the results obtained in the three semi-leptonic channels.}
  \label{tab:res189gw}
 \end{center}
 \end{table}


\begin{thebibliography}{99}

\bibitem{delpaper161} DELPHI Collaboration, P.~Abreu {\it et al.}, 
        \Journal{\PLB}{397}{158}{1997} 
\bibitem{delpaper172} DELPHI Collaboration, P.~Abreu {\it et al.},
        \Journal{\EUR}{2}{581}{1998}
\bibitem{delpaper183} DELPHI Collaboration, P.~Abreu {\it et al.},
        \Journal{\PLB}{462}{410}{1999}
\bibitem{lepmw} ALEPH Collaboration, R.~Barate {\it et al.},
\Journal{\EUR}{C17}{241}{2000} \\
                L3 Collaboration, M.~Acciarri {\it et al.},
        \Journal{\PLB}{454}{386}{1999}\\
                OPAL Collaboration, G.~Abbiendi {\it et al.},
        hep-ex/0009018 (2000) 
\bibitem{hadmw} CDF Collaboration, T.~Affolder {\it et al.},
        hep-ex/0007044 (2000), to be published in {\PRD} \\
                D0 Collaboration, B.~Abbott {\it et al.},
                \Journal{\PRL}{80}{3008}{1998} \\
                D0 Collaboration, B.~Abbott {\it et al.},
        \Journal{\PRL}{84}{222}{2000}

\bibitem{cdfwidth} CDF Collaboration, T.~Affolder {\it et al.},
        \Journal{\PRL}{85}{3347}{2000}

\bibitem{delphi} DELPHI Collaboration, P.~Aarnio {\it et al.},
        \Journal{\NIMA}{303}{233}{1991} \\
                 DELPHI Collaboration, P.~Abreu {\it et al.}, 
        \Journal{\NIMA}{378}{57}{1996} 
\bibitem{excal} F.A. Berends, R. Pittau, R. Kleiss, 
        \Journal{\CPC}{85}{437}{1995} 

\bibitem{qedps} Y. Kurihara, J. Fujimoto, T. Munehisha, Y. Shimizu,  
        Progress of Theoretical Physics Vol. {\bf 96} (1996) 1223  

\bibitem{pythia} T. Sj\"ostrand, 
        \Journal{\CPC}{82}{74}{1994}

\bibitem{twogam} S. Nova, A. Olchevski and T. Todorov {\it } in {\it Physics at LEP2}, CERN 96-01 Vol.2 (1996) 224

\bibitem{deltune} DELPHI Collaboration, P.~Abreu {\it et al.},
        \Journal{\ZPC}{73}{11}{1996}

\bibitem{koralw} S. Jadach, W. Placzek, M. Skrzypek,B.F.L. Ward, Z. Was,
        \Journal{\CPC}{119}{272}{1999}

\bibitem{LUCLUS} T. Sj\"ostrand,
        {\it PYTHIA 5.7 and JETSET 7.4: Physics and manual}, 
        CERN-TH-7112-93-REV (1995)

\bibitem{neural} Code kindly provided by J. Schwindling and B. Mansoulie

\bibitem{189xsec} DELPHI Collaboration, P.~Abreu {\it et al.},
         \Journal{\PLB}{479}{89}{2000}

\bibitem{SPRIM} P. Abreu et al., 
        \Journal{\NIMA}{427}{487}{1999} 

\bibitem{DURHAM} S. Catani, Yu.L. Dokshitzer, M. Olsson, G. Turnock, B.R. Webber,
        \Journal{\PLB}{269}{432}{1991} \\
        N. Brown, W. Stirling, 
        \Journal{\ZPC}{53}{629}{1992} 

\bibitem{aabtag} G. Borisov, 
        \Journal{\NIMA}{417}{384}{1998} \\
                 DELPHI Collaboration, P.~Abreu {\it et al.}, 
        \Journal{\EUR}{10}{415}{1999}

\bibitem{CAMJET} Yu.L. Dokshitzer, G.D. Leder, S. Moretti, B.R. Webber,
        JHEP 08 (1997) 001

\bibitem{DICLUS} L. L\"onnblad, 
        \Journal{\ZPC}{58}{471}{1993} 

\bibitem{rudolph} G.Rudolph, 
{\it ALEPH \HERWIG\ tuning} - private communication

\bibitem{mlbz} N.Kjaer, M.Mulders, 
         CERN-OPEN-2001-026 (2001)


\bibitem{energy} LEP Energy Working Group, 
                 \Journal{\EUR}{11}{573}{1999} \\
                 LEP Energy Working Group, 
                 LEP Energy working group note 99/01 (1999)

\bibitem{bose} 
         A. Valassi, {\it Bose-Einstein correlations in W decays}, hep-ex/0009039 (2000), to be published in the proceedings of the 30th International Conference On High-Energy Physics, 27 Jul - 2 Aug 2000, Osaka, Japan 

\bibitem{luboei} L. L\"onnblad, T. Sj\"ostrand
        \Journal{\EUR}{2}{165}{1998} 

\bibitem{kkm} V. Kartvelishvili, R. Kvatadze, R. M{\o}ller,
        \Journal{\PLB}{408}{331}{1997}

\bibitem{sharka} S. Todorova, J. Rames,
        IReS-97-29 PRA-HEP-97-16, hep-ph/9710280 (1997) 
        
\bibitem{L3Colour} 
         P. de Jong, {\it Color reconnection in W decays}, hep-ex/0103018 (2001), to be published in the proceedings of the 30th International Conference On High-Energy Physics, 27 Jul - 2 Aug 2000, Osaka, Japan

\bibitem{opalariadne} OPAL Collaboration, G.~Abbiendi {\it et al.},
        \Journal{\EUR}{11}{217}{1999}

\bibitem{perturb} T. Sjostrand, V. Khoze,
        \Journal{\ZPC}{62}{281}{1994}\\
        T. Sjostrand, V. Khoze,
        \Journal{\PRL}{72}{28}{1994}

\end{thebibliography}
\end{document}